\documentclass[format=acmsmall,screen=false,review=false]{acmart}

\usepackage{amsmath}
\usepackage{graphicx}
\usepackage{textcomp}
\usepackage{xcolor}
\usepackage{balance}
\usepackage{float}
\usepackage{import}
\usepackage{booktabs} 
\usepackage{import}
\usepackage{graphicx}
\usepackage{subcaption}
\usepackage{caption}
\usepackage[utf8]{inputenc}
\usepackage[export]{adjustbox}
\usepackage{wrapfig}
\usepackage{algorithm}
\usepackage[noend]{algpseudocode}

\usepackage{multirow}
\usepackage{enumitem}
\usepackage{varwidth}
\usepackage{comment}
\usepackage{multicol}

\AtBeginDocument{%
  \providecommand\BibTeX{{%
    \normalfont B\kern-0.5em{\scshape i\kern-0.25em b}\kern-0.8em\TeX}}}

\begin{document}

\title{Breaking On-Chip Communication Anonymity using Flow Correlation Attacks}

\author{Hansika Weerasena}
\affiliation{%
  \institution{University of Florida}
  \city{Gainesville}
  \state{FL}
  \postcode{32611}
  \country{USA}
}
\email{hansikam.lokukat@ufl.edu}

\author{Prabhat Mishra}
\affiliation{%
  \institution{University of Florida}
  \city{Gainesville}
  \state{FL}
  \postcode{32611}
  \country{USA}
}
\email{prabhat@ufl.edu}

\begin{abstract}
  Network-on-Chip (NoC) is widely used to facilitate communication between components in sophisticated System-on-Chip (SoC) designs. Security of the on-chip communication is crucial because exploiting any vulnerability in shared NoC would be a goldmine for an attacker that puts the entire computing infrastructure at risk. We investigate the security strength of existing anonymous routing protocols in NoC architectures, making two pivotal contributions. Firstly, we develop and perform a machine learning (ML)-based flow correlation attack on existing anonymous routing techniques in Network-on-Chip (NoC) systems, revealing that they provide only packet-level anonymity. Secondly, we propose a novel, lightweight anonymous routing protocol featuring outbound traffic tunneling and traffic obfuscation. This protocol is designed to provide robust defense against ML-based flow correlation attacks, ensuring both packet-level and flow-level anonymity. Experimental evaluation using both real and synthetic traffic demonstrates that our proposed attack successfully deanonymizes state-of-the-art anonymous routing in NoC architectures with high accuracy (up to 99\%) for diverse traffic patterns. It also reveals that our lightweight anonymous routing protocol can defend against ML-based attacks with minor hardware and performance overhead.
\end{abstract}





\settopmatter{printacmref=false} 
\renewcommand\footnotetextcopyrightpermission[1]{} 
\pagestyle{plain} 

\maketitle

\section{Introduction}
\label{sec:introduction}

Advanced manufacturing technology allows the integration of heterogeneous Intellectual Property (IP) cores on a single System-on-Chip (SoC). For example, Intel's Xeon® Scalable Processor~\cite{Intelxeon} supports up to 64 cores. 
Traditional bus architectures fail to scale up with the communication requirements of the increasing number of IP cores. Network-on-Chip (NoC) is the preferred communication fabric to meet the high throughput and scalability requirements between these IP cores. Due to time to market constraints and cost-effectiveness, SoC manufacturers tend to use third-party vendors and services from the global supply chain~\cite{mishra2021network}. Typically only a few IP cores are designed in-house, while others are reusable IPs from third-party vendors. For example, FlexNoc interconnect is used by four out of the top five fabless companies to facilitate their on-chip communication \cite{js2015runtime}. A long and potentially untrusted supply chain can lead to the introduction of malicious implants through various avenues, such as untrusted  CAD  tools, rogue designers, or at the foundry. Furthermore, these sophisticated SoC designs make it harder to do complete security verification~\cite{mishra2017hardware}. While designing energy-efficient NoCs is a primary goal today, securing them is equally crucial as exploiting an NoC could allow attackers to access communications between IP cores and compromise the entire computing infrastructure's security.

\begin{figure}[t]
\centering
\includegraphics[width=0.6\linewidth]{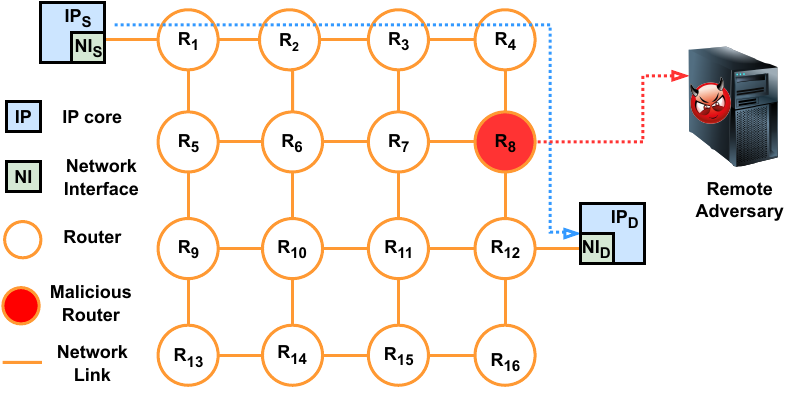}
\vspace{-0.15in}
\caption{In a 4x4 mesh NoC, each IP connects to NoC via a network interface and router. A malicious router can intercept packets between $IP_S$ and $IP_D$, forwarding them to a remote adversary for sophisticated attacks.}
\label{typical_noC_security_threat_model}
\vspace{-0.2in}
\end{figure}

Figure~\ref{typical_noC_security_threat_model} shows a $4\times4$ mesh NoC where mesh topology is the most commonly used topology in NoC. A single tile consists of an IP core, Network Interface (NI), and Router. Security issues in a typical NoC can be classified based on various security goals (confidentiality, integrity, anonymity, authenticity, availability, and freshness) compromised by an attacker~\cite{weerasena2023security}. There are efficient detection and mitigation of security vulnerabilities~\cite{charles2020lightweight,ancajas2014fort,sepulveda2017towards, farahmandi2019system, charles2020securing, ganguly2011complex} for securing NoC-based SoCs. In a typical NoC, to enable fast packet forwarding, the header information is kept as plaintext while the packet data is encrypted. An adversary can implant a hardware Trojan in a router ($R_8$ in Figure \ref{typical_noC_security_threat_model}), which can collect packets from the same source-destination pair and send them to a remote adversary that can launch traffic and metadata analysis attacks~\cite{weerasena2023security}. For example, imagine a source node ($IP_S$) is a cryptographic accelerator that needs to communicate with a memory controller, destination node ($IP_D$), to facilitate memory requests for the cryptographic operation. An adversary can use a malicious router in the middle to collect packets between $IP_S$ and $IP_D$ over a time interval and recover the key by launching a ciphertext-only cryptanalysis attack~\cite{charles2020lightweight,sarihi2021securing,patooghy2023securing}. Similarly, a collection of packets belonging to the same communication session can also be analyzed to discover what program is running at $IP_S$ or reverse engineer the architectural design using a simple hardware Trojan and powerful remote adversary~\cite{ahmed2020defense,ahmed2021can,dhavlle2023defense}. Ensuring anonymity in NoC communication can mitigate metadata and traffic analysis attacks since anonymity ensures that there is no unauthorized disclosure of information about communicating parties. Recent literature features two anonymous routing approaches for securing NoC traffic: ARNoC~\cite{charles2020lightweight} and a stochastic anonymous routing (SAR) protocol~\cite{sarihi2021securing, patooghy2023securing}. Although these anonymous routing solutions provide packet-level anonymity, we show that they fail to provide flow-level anonymity by breaking anonymity via flow correlation attacks. After breaking anonymity, the adversary can launch various traffic and metadata analysis attacks on the deanonymized communication session. Specifically, this paper evaluates the security strength of anonymous routing protocols in NoCs and makes the following major contributions. 

    \begin{itemize}
        \item We propose an attack on existing anonymous routing by correlating NoC traffic flows via machine learning (ML).
        \item We show that our ML-based attack can break the anonymity of the state-of-the-art anonymous routing (ARNoC~\cite{charles2020lightweight} and SAR~\cite{sarihi2021securing, patooghy2023securing}) and validates the need for flow-level anonymity. 
        \item The robustness of the attack is assessed across diverse configurations and traffic patterns.
        \item We propose a novel anonymous routing protocol with outbound traffic tunneling and obfuscation as a lightweight countermeasure that ensures packet-level and flow-level anonymity.
        \item Experimental results demonstrate that our countermeasure can defend against flow correlation attacks with minor hardware and performance overhead.
    \end{itemize}

The remainder of this paper is organized as follows: Section~\ref{sec:background_and_related_work} provides relevant background and surveys related efforts. Section~\ref{sec:attack} describes our ML-based attack on anonymous routing. Section~\ref{sec:countermeasure} proposes a lightweight protocol to defend against such attacks. Section~\ref{sec:exp-eval} presents experimental results and evaluation. Finally, the paper is concluded in Section~\ref{sec:conclusion}.

\section{Background and Related Work}\label{sec:background_and_related_work}

This section provides the relevant background and surveys the related efforts to highlight the novelty of this work.

\vspace{-0.15in}
\subsection{Network-on-Chip (NoC) Traffic}

NoC enables communication by routing packets through a series of nodes. There are two types of packets that are injected into the network: control and data packets. Consider an example when a processor ($IP_S$) wants to load data from a particular memory controller ($IP_D$), it will issue a control packet requesting the data from memory. The packet travels through routers following a predefined routing protocol. Upon reaching the destination IP, it responds with a data packet containing the requested data. In general, header information is kept as plaintext and the payload data is encrypted. At each source NI, the packets are divided into fixed-size flits, which is the smallest unit used for flow control. There is a head flit followed by multiple body flits and tail flits. Routing in NoC can be either deterministic or adaptive; both approaches use header information to make routing decisions at each router. XY routing is the most commonly used routing in mesh-based traditional NoCs, which basically takes all the X links first, followed by Y links. NoC uses links to connect different components of the interconnects. Links can be either internal or boundary. A boundary link is a link that connects a router to a network interface, while internal links connect two routers. Our ML-based attack on anonymous routing makes use of the flow of flits (inter-flit delays), whereas our countermeasure manipulates routing decisions to create virtual tunnels.

\vspace{-0.1in}
\subsection{Attacks on Anonymity and Anonymous Routing}
\label{subsec:anonymous_routing}

In the context of communication, anonymity refers to the quality of being unidentifiable within a set of subjects. The primary goal of anonymity is to protect the privacy of communicating parties. Traffic and metadata analysis are two types of attacks that compromise the lack of anonymity in NoC communication~\cite{weerasena2023security}. A traffic analysis attack collects packets in a particular communication session between parties and analyzes them to deduce various aspects, such as what type of application is running in an SoC. Similarly, metadata analysis attacks use ancillary data of communications, such as sender and receiver information, time stamps, and packet sizes, to compromise the privacy of communication parties. Anonymous routing hides the identity of the communicating parties from anyone listening in the middle and hinders the effectiveness of these attacks. In this context, we consider two types of anonymity: packet-level and flow-level anonymity. Packet-level anonymity focuses on concealing individual data packets' origin, destination, and content, while flow-level anonymity aims to obscure the relationship between packets in a communication session. 

The Tor network~\cite{dingledine2004tor}, based on onion routing, and the I2P network~\cite{zantout2011i2p}, based on garlic routing, are key examples of anonymous routing in traditional networks. Onion routing creates tunnels through multiple hops, encrypting the message in layers equal to the number of hops. Each hop peels off a layer to gradually reveal the original message. Garlic routing extends onion routing by bundling and encrypting multiple messages together, similar to garlic cloves. 
Many attacks target Tor network anonymity, such as the flow correlation attack~\cite{nasr2018deepcorr}, but they cannot be directly applied to NoC for three key reasons. (1) Traffic characteristics differ significantly between NoC and traditional network due to varying use cases. NoC is used for routing simple on-chip communication traffic, such as cache coherence, memory accesses, and inter-processor communications. In contrast, traditional networks handle complex use cases, such as enterprise data services, cloud computing tasks, and multimedia streaming. (2) The existing attack relies heavily on packet size as a feature, whereas NoC flits are the fundamental unit of flow control, and they are of fixed size. (3) In NoCs, all nodes function as onion routers, unlike traditional networks which mix normal and onion routers.




\subsection{Related Work}
\label{subsec:related_work_s}

The security of on-chip communication has been extensively studied, encompassing a wide range of attacks and countermeasures, including eavesdropping attacks~\cite{ancajas2014fort, raparti2019lightweight, weerasena2021lightweight, sepulveda2017towards, charles2020securing}, spoofing attacks~\cite{ancajas2014fort, weerasena2024lightweight,lebiednik2018spoofing}, denial of service (DoS) attacks~\cite{boraten2016packet, sinha2021sniffer, sudusinghe2021denial, manju2020sectar,js2015runtime}, side-channel attacks~\cite{dai2022don,boraten2018securing,weerasena2023revealing, reinbrecht2016gossip}, and packet tampering attacks~\cite{charles2020lightweight, sepulveda2017towards, wang2020tsa}. Anonymity is crucial for secure on-chip communication, but solutions in the traditional networks are too expensive for resource-constrained NoCs.
An anonymous routing protocol (SAR) that needs NoC packets to be identified as secure and non-secure packets is presented in~\cite{patooghy2023securing,sarihi2021securing}. This approach stochastically selects a routing scenario for each packet out of three scenarios available to confuse adversaries. ~\citeauthor{charles2020lightweight}~\cite{charles2020lightweight} presented an anonymous routing solution (ARNoC) for NoC based on onion routing~\cite{dingledine2004tor} to ensure the anonymity of a communication session. ARNoC creates an on-demand anonymous tunnel from the source to the destination where intermediate nodes know only about the preceding and succeeding nodes. Our proposed ML-based attack can break the anonymity of both ARNoC and SAR. 


A threat model based on the insertion of Hardware Trojans (HTs) in network links is addressed in ~\cite{yu2013exploiting, boraten2016mitigation}. \citeauthor{yu2013exploiting} \cite{yu2013exploiting} show that the Trojans can be inserted in boundary links and center links that can do bit flips in the header packet that can lead to deadlock, livelock, and packet loss. ~\citeauthor{boraten2016mitigation}~\cite{boraten2016mitigation} discuss the DoS attacks that can be launched by malicious links. This specific Trojan performs packet injection faults at the links, triggering re-transmissions from the error-correcting mechanism. \citeauthor{ahmed2021can}~\cite{ahmed2021can} introduce the concept of Remote Access Hardware Trojan (RAHT), where a simple HT in NoC can leak sensitive information to an external adversary who can launch complex traffic analysis attacks. These RAHTs are hard to detect due to negligible area, power, and timing footprint. Recent efforts~\cite{ahmed2020defense,dhavlle2023defense} utilize a similar threat model that can reverse engineer applications through traffic analysis attacks. A threat model where an HT in NoC collaborates with a colluding application is used to launch multitudes of attacks in the NoC literature~\cite{charles2020lightweight,raparti2019lightweight,hussain2018eetd,jyv2018run,sepulveda2017towards,boraten2016packet}. Our proposed attack assumes malicious boundary links as the points of data collection that gets remote access to external adversary through a colluding application.

ML-based techniques have been used to detect and mitigate attacks on NoCs in \cite{sudusinghe2021denial, wang2020tsa, sinha2021sniffer}. \citeauthor{sudusinghe2021denial}~\cite{sudusinghe2021denial} used several ML techniques to detect DoS attacks on NoC traffic. Reinforcement learning is used by \cite{wang2020tsa} to detect HTs in NoC at run time. 
~\citeauthor{sinha2021sniffer}~\cite{sinha2021sniffer} use an ML-based approach to localize flooding-based DoS attacks. 
None of these approaches consider attacks or countermeasures related to anonymous routing in NoC architectures.  {\it To the best of our knowledge, our study is the first attempt to deanonymize exiting anonymous routing protocols via ML-based flow correlation attack and propose a lightweight countermeasure with packet-level and flow-level anonymity for NoC-based SoCs.}


\vspace{-0.1in}
\subsection{Flow Correlation Challenges}

NoC traffic flow can be considered a time series data array with values of increasing timestamps in order.
For example, in a communication session, we can consider an array of time differences between each packet coming into a node as a flow. Flow correlation is when we take two such pairs and compare if they are correlated in some manner. For example, in a network link, the flow of inter-flit delay entering and going out of the link are correlated. Though correlating outgoing and incoming traffic on a link seems straightforward, correlating traffic between two nodes in a large network with multiple hops in NoC is extremely difficult for the following reasons:

\begin{itemize}
    \item Queuing delay at each hop is unpredictable and can interfere with traffic flow characteristics.
    \item A pair of correlated nodes may communicate with other nodes, which is considered as noise.
    \item The communication path of the correlated pair may be shared by other nodes in SoC, which will interfere with the traffic flow characteristics between correlated pairs.

    
\end{itemize}

\section{ML-based Attack on Anonymous Routing}
\label{sec:attack}

\begin{figure}[t]
\includegraphics[width=0.6\linewidth]{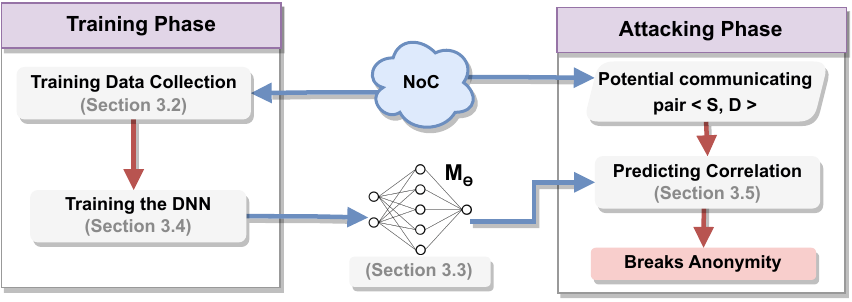}
\vspace{-0.15in}
\caption{ Overview of our proposed ML-based attack that consists of two phases (training and attacking). }
\label{fig:attack_overview}
\end{figure}

We first outline the threat model used in the proposed attack. Next, we describe our data collection, training, and application of the ML model to accomplish the attack.

\vspace{-0.1in}
\subsection{Threat Model}
\label{subsec:threat_model}

The threat model considers an NoC that uses encrypted traffic and anonymous routing, either ARNoC~\cite{charles2020lightweight} or SAR~\cite{sarihi2021securing,patooghy2023securing}. Thus, we consider traffic to have packet-level anonymity; attackers cannot identify the sender/receiver due to anonymous routing. Furthermore, they cannot recover the payload due to encryption. The threat model consists of three major components: (1) a malicious NoC, (2) a malicious program (\textit{collector}), and (3) a pre-trained ML model.

\vspace{0.05in}
\noindent \underline{\it Malicious NoC}: The malicious NoC has malicious boundary links with Hardware Trojan (HT). The HT counts the number of cycles between incoming and outgoing flits (inter-flit delay) to and from an IP. After specific intervals, HT gathers all inter-flit delay into an array and sends it to the IP where the malicious program (\textit{collector}) is running. HT can be inserted by various adversaries in the extended supply chain, such as through untrusted CAD tools, rogue engineers, or at the foundry via reverse engineering, and remain undetected during post-silicon verification~\cite{mishra2017hardware}. A similar threat model of inserting HT at NoC links has been discussed in~\cite{yu2013exploiting,boraten2016mitigation}. Note that the area and power overhead of an HT is negligible in a large MPSoC~\cite{ahmed2020defense}.

\vspace{0.05in}
\noindent \underline{\it Malicious Program}: Cloud infrastructures use multi-core SoCs in multi-tenant
platforms where they are virtualized and allocated to various applications from different users.
The attacker disguised as one of the multiple users
of this shared virtualized system can easily launch a malicious program and stay undetected. The \textit{collector} is such a malicious program; it activates/deactivates HT to keep it hidden from any run-time HT detection mechanisms. The main functionality of the \textit{collector} is to collect inter-flit-delays from HT-infected links and send them to the ML model. This threat model, where a malicious NoC with an HT collaborates with a colluding application in same SoC ( i.e. \textit{collector}), is a well-documented approach in NoC security literature~\cite{weerasena2023security, charles2021survey}.

\begin{figure}[t]
\includegraphics[width=0.55\linewidth]{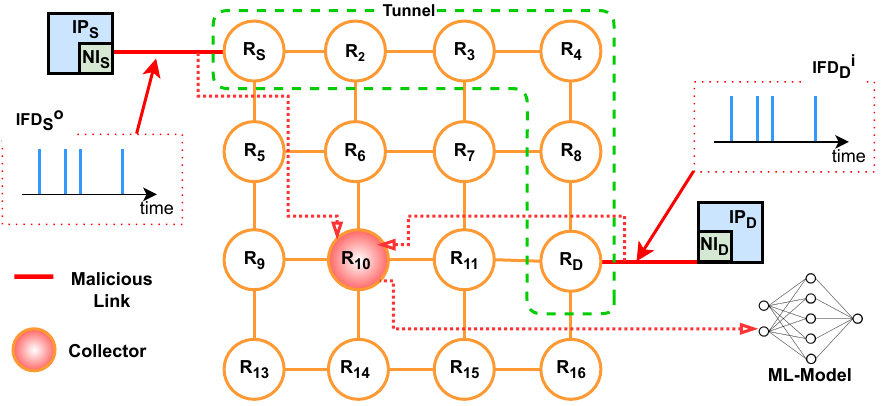}
\vspace{-0.1in}
\caption{Malicious boundary links outside the anonymous tunnel extract flow pair ($IFD_S^o$, $IFD_D^i$) and send them to the collector. Then, \textit{collector} sends them to ML-model.}
\label{fig:threat_model_for_attack_on_onion_routing}
\vspace{-0.1in}
\end{figure}

\vspace{0.05in}
\noindent \underline{\it ML Model}: The pre-trained ML model runs in a remote server/cloud controlled by the adversary. The flow correlation uses the attacking phase out of two phases (training and attacking) of the ML model. The training phase is detailed in Section \ref{subsec:train_DNN}. The attacking phase classifies whether two inter-flit delay arrays are correlated or not.  Figure~\ref{fig:attack_overview} shows a high-level overview of the proposed flow correlation attack from the perspective of the ML model. The training phase is performed offline and is responsible for collecting training data and training of the ML model. The adversary can use an emulator or simulator mimicking the target system to collect data. The adversary can generate a large amount of trained data by changing process mapping, benchmarks, and other traffic characteristics (as discussed in Section~\ref{sec:ex-data-col}) to make the model generic. While training the ML model for detecting correlation can be computationally expensive, it is not a limiting factor since the training is a one-time activity. Note that the model can be retrained, if needed after specific intervals, to ensure that it remains effective and up-to-date throughout its operational lifetime.




Figure~\ref{fig:threat_model_for_attack_on_onion_routing} shows an example of the attacking phase on ARNoC. In ARNoC, a tunnel exists between source and destination routers if their associated IPs are in a communication session. ARNoC forms the tunnel to ensure anonymity by hiding the headers. The HTs in the links are in the inactive state by default. The \textit{collector} periodically checks the state of all infected boundary links and flags communicating links as suspicious. This is done via monitoring a simple heuristic of inbound/outbound packet counts between two nodes. The collector will examine these counts and instruct the HT to start collecting inter-flit delays if the difference is within a specified threshold.  Imagine a scenario where an adversary suspects communication between the source ($IP_S$) and destination ($IP_D$); the \textit{collector} activates HT associated with the boundary links of $IP_S$ and $IP_D$. On activation, HTs start sending periodic inter-flit delay arrays to the collector. More specifically, the Trojan will observe and leak both outbound ($IFD_S^o$) and inbound ($IFD_D^i$) traffic flows. Here, $IFD_S^o$ refers to the outbound inter-flit delay arrays from the source IP, and  $IFD_D^i$ refers to the inbound inter-flit delay arrays at the destination IP. Upon receiving inter-flit delay arrays, the \textit{collector} is responsible for sending collected data on inter-flit delay to the ML model. The adversary uses the ML model to pinpoint two specific nodes that are communicating and breaks the anonymity. 

After breaking anonymity through proposed flow correlation, an attacker can launch either metadata or traffic analysis attacks~\cite{ahmed2021can,ahmed2020defense,dhavlle2023defense,charles2020lightweighta,sarihi2021securing}, as discussed in Section~\ref{sec:introduction} and ~\ref{sec:background_and_related_work}. Breaking anonymity can have significant consequences in scenarios where preserving the anonymity of data traffic is critical. For example, in the case of confidential computing~\cite{costan2016intel}, it can leak the host memory region of an application by breaking anonymity between the computing node and the memory controller. Furthermore, after breaking anonymity, attackers can use it as a stepping stone for more advanced attacks, such as targeted denial-of-service attacks.

\begin{figure}[t]
\centering
\includegraphics[scale =0.30]{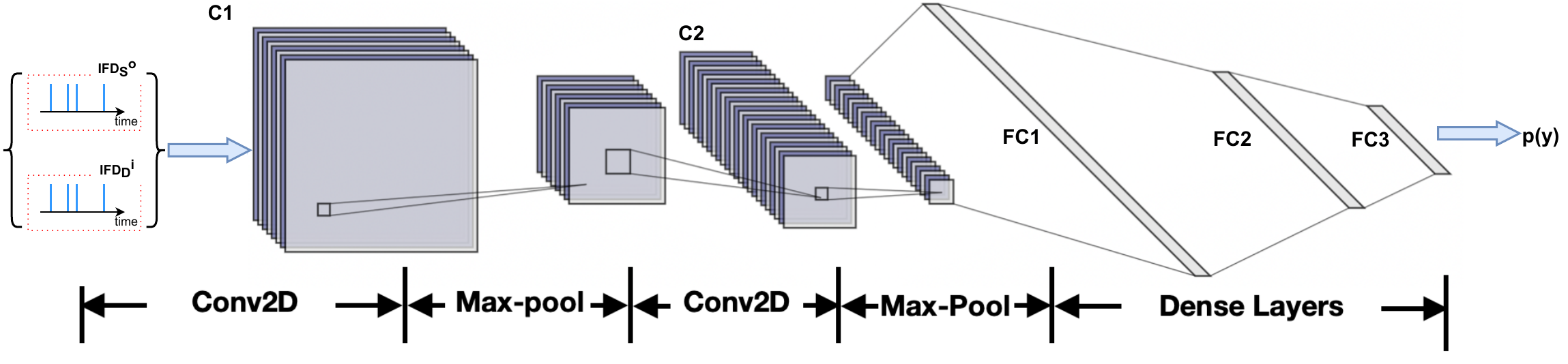}
\vspace{-0.2in}
\caption{DNN architecture has two convolution layers (C1, C2) and three fully connected layers (FC1 - FC3).}
\label{fig:dnn}
\vspace{-0.1in}
\end{figure}

\subsection{Collecting Data for Training}
\label{subsec:data_collect}

Algorithm \ref{algo:data_collect} outlines the training data collection when running ARNoC or SAR. We collect inbound and outbound inter-flit delays for all source and destination IPs (line \ref{algo:data_collect_line5}). Then, we label each flow pair as either `1' or `0' according to the  ground truth (line \ref{algo:data_collect_line6}). If $IP_S$ and $IP_D$ of flow pair $\{IFD_S^o, IFD_D^i\}$ are correlated to each other ($IP_S$ and $IP_D$ communicating in a session), the flow pair is tagged as `1' and otherwise `0'. These tagged flow pairs are utilized as the training set. Note that only the first $l$ \text  elements of each flow of flow pair ($\{IFD_S^o, IFD_D^i\}$) will be used in the training and testing. We model external traffic interference on correlated flows by considering two scenarios: other nodes communicating with the correlated pair and with each other, reflecting shared resource and resource path. We use a deep neural network (DNN) as the ML model for our proposed flow correlation attack. 
To ensure a generic dataset and sufficient data for DNN training, we conduct multiple iterations of data collection (Algorithm~\ref{algo:data_collect}), varying the mapping of correlated pairs to different NoC nodes each time. Section \ref{sec:exp-eval} elaborates on synthetic and real traffic data collection.

\begin{algorithm}[h]
    \caption{Data Collection}
    \label{algo:data_collect}
    \begin{algorithmic}[1]
        \State X, Y $ \gets \emptyset$ \label{algo:data_collect_line2}
        \Procedure{CollectData ()}{} 
                \For{$\forall$ $(s,$ $d)$ $\in$  $(S$, $D)$}
                    \State $X$ $\leftarrow$ $X$ $\cup$ $\{$ $IFD_s^o$, $IFD_d^i$ $\}$ \label{algo:data_collect_line5}
                    \State $Y$ $\leftarrow$ $Y$ $\cup$ $c$ $\colon$ $c$ $\in$ $\{$ $0$, $1$ $\}$ \label{algo:data_collect_line6}
                \EndFor
          \State \textbf{return} $X$, $Y$
          \EndProcedure
    \end{algorithmic}
\end{algorithm}

\subsection{DNN Architecture}
\label{subsec:dnn-arc}

We carefully examined various configurations and reached out to the final DNN architecture shown in Figure~\ref{fig:dnn}. We selected Convolution Neural Networks (CNN)~\cite{schmidhuber2015deep} as our model architecture for the following reasons. First, since multivariate time series have the same 2-dimensional data structures as images, CNN for analyzing images is suitable for handling multivariate time series~\cite{zhao2017convolutional}. Second, recently published works using CNN for flow correlation~\cite{guo2019deep, nasr2018deepcorr} has shown promising results. Our final architecture has two convolution layers followed by three fully connected layers to achieve promising performance. The first convolution layer (C1) has $k_1$ number of kernels of size (2, $w_1$). The second convolution layer (C2) has $k_2$ number of kernels of size (2, $w_2$). The main intuition of C1 is to identify and extract the relationship between two traffic flows ($IFD_S^o$, $IFD_D^i$), while we assign the task of advancing features to C2. In our approach, both C1 and C2 have a stride of (2, 1). A max-pooling layer immediately follows both convolution layers. Max pooling uses a max operation to reduce the dimension of features, which also logically reduces overfitting. Finally, the result of C2 is flattened and fed to a fully connected network with three layers. Additionally, the set ($k_1$, $k_2$, $w_1$, $w_2$) are considered as hyper-parameters. We provide details on hyper-parameter tuning in Section~\ref{sec:hyper}.  We use ReLU as the activation function for all convolution and fully connected layers to avoid the vanishing gradient problem and improve performance. Due to the fact that our task is a binary classification, we apply a \textit{sigmoid} function in the last output layer to produce predictions.

\subsection{Training the DNN Model}
\label{subsec:train_DNN}

Algorithm \ref{algo:train} outlines the major steps in the training process of the ML model. Specific sizes and parameters used in training are outlined in Section~\ref{sec:exp-eval}. We train the DNN over multiple epochs (line \ref{alg_line:train-line10}) using labeled inter-flit delay distributions as the input. During the training phase, the stochastic gradient descent (\textit{sgd})  optimizer minimizes the loss and updates the weights in the DNN (line \ref{alg_line:train-line14}).
To achieve this binary classification results from the last fully connected layer pass through a \textit{sigmoid} layer~\cite{han1995influence} (line \ref{alg_line:train-line12}) to produce classification labels.  

\begin{algorithm}[h]
  \caption{ML Model Training }
  \label{algo:train}
  \begin{algorithmic}[1]
  \State $X$ : [$x_1$, ..., $x_j$, ..., $x_N$] where $x_j =\{$ $IFD_s^o$, $IFD_d^i$ $\}_j$
  \State $Y$ : [$y_1$, ..., $y_j$, ..., $y_N$] where $y_j \in \{ 0,1\}$
  \Procedure{TrainModel }{$X$, $Y$} 
        \State $Circuit$ $samples$ $X$ $and$ $labels$ $Y$
        \State $Model$ $M_{\Theta}$ $initialization$ 
    \For{$epoch$ $\in$ [$1$, ..., $NoOfEpochs$]}
    \label{alg_line:train-line10}
    \For{$x_j \in X$ and $y_j \in Y$}
        \State $out_j$ = $sigmoid$( $M_{\Theta}(x_j)$ )
        \label{alg_line:train-line12}
        \State $loss$ = $\sum\limits^N_j$ cross\_entropy$(out_j, y_j)$
        \label{alg_line:train-line13}
        \State $\Theta$ = sgd($\Theta, \nabla loss$)
        \label{alg_line:train-line14}
    \EndFor
    \EndFor
    \State Return $M_{\Theta}$
    \label{alg_line:train-line15}
  \EndProcedure
  \end{algorithmic}
\end{algorithm}

Formally, the \textit{sigmoid} layer is a normalized exponential function
$
f(x) = \frac{1}{1+ e^{-x}}
$, 
which aims at mapping the given vector to a probability value that lies in $[0,1]$. 
The value of the output of the last layer is the predicted label $p(y)$ which can be denoted as:
\vspace{-0.05in}
\begin{align*}
    p(y) = \frac{1}{1+e^{-(M(s,d))}}
    \vspace{-0.1in}
\end{align*}
where $s$ and $d$ denote the source and destination input distribution respectively, and $M$ denotes a function map for the entire DNN model. Since it is a binary classification task, for given input $(s,d)$ pairs' labels, their probability distributions are either $(1,0)$ for `true' (correlated) and $(0,1)$ for `false' (uncorrelated). Therefore, we choose \textit{binary cross-entropy} (line \ref{alg_line:train-line13}) as the loss function as follows: 
\vspace{-0.2in}
\begin{align*}
loss(p(y)) = -\frac{1}{N} \sum\limits^{N}_{i=1} y_i\cdot log(p(y_i)) +  (1-y_i)\cdot log(1-p(y_i))
\vspace{-0.1in}
\end{align*}
where $y$ is the label (1 for correlated pairs and 0 for uncorrelated pairs), and $N$ is the total number of training samples. The goal of model training is to minimize the loss function by gradient descent for multiple iterations, where in each step the model parameters $\Theta$ are updated by 
$
\Theta' = \Theta  + \nabla loss(p(y))
$.

\vspace{-0.1in}
\subsection{Predicting Correlation}
\label{subsec:breaking_anonimity}

The trained model is used in attacking phase as shown in Algorithm \ref{algo:attack}. During the attacking phase, we feed the two inter-flit delay arrays from a suspicious source ($S$) and destination ($D$) of the ongoing communication session to the ML model (lines 4-5). The ML model will output 1 if the source and destination are communicating, and 0 otherwise (lines 5). If $S$ and $D$ are communicating and the ML model output is 1, our attack has successfully broken the anonymity.

\begin{algorithm}[htp]
  \caption{Attack on Anonymous Routing}
  \label{algo:attack}
  \begin{algorithmic}[1]
  \State $IFD_S^o$ : outbound inter-flit delay array of S
  \State $IFD_D^i$ : inbound inter-flit delay array of D
  \State $M_{\Theta}$ : pre-trained model
  \Procedure{Attack }{ $\{IFD_S^o, IFD_D^i\}$, $M_{\Theta}$} 
    \label{alg_line:attack-line2}
    \State $p(y)$ $\gets$ \textit{ predict($\{ IFD_S^o ,IFD_D^i\}$, $M_{\Theta}$)} 
    \State \Return $p(y)$
    \label{alg_line:attack-line15}
  \EndProcedure
  \end{algorithmic}
\end{algorithm}
\vspace{-0.1in}
\section{Defending against ML-based Attacks}
\label{sec:countermeasure}

\begin{figure}[tp]
\includegraphics[width=\linewidth]{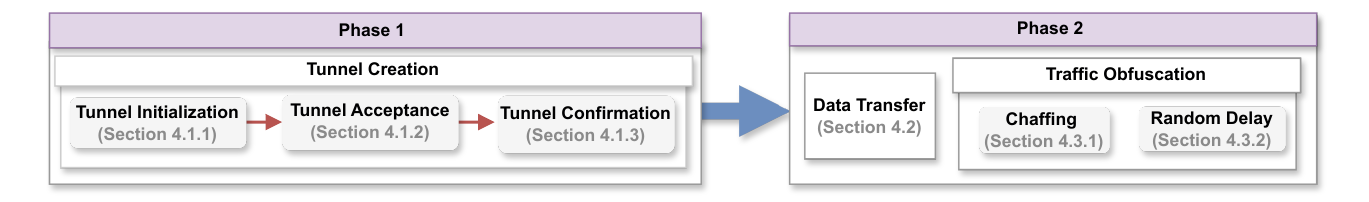}
\vspace{-0.3in}
\caption{ Overview of the proposed lightweight anonymous routing to defend against flow correlation attack. It has two phases: tunnel creation and data transfer with traffic obfuscation. }
\vspace{-0.1in}
\label{fig:overview_of_counter}
\end{figure}

In this section, we propose a novel lightweight anonymous routing protocol as a countermeasure against the ML-based attack described in Section~\ref{sec:attack}. Figure \ref{fig:overview_of_counter} shows an overview of our  proposed anonymous routing that consists of two phases: 1) outbound tunnel creation and 2) data transfer with traffic obfuscation. We utilize two obfuscation techniques (chaffing of flits and random delays).


\vspace{-0.1in}
\subsection{Outbound Tunnel Creation}
\label{subsec:outbound_tunnel_creation}



\begin{figure}[tp]
\includegraphics[width=0.55\linewidth]{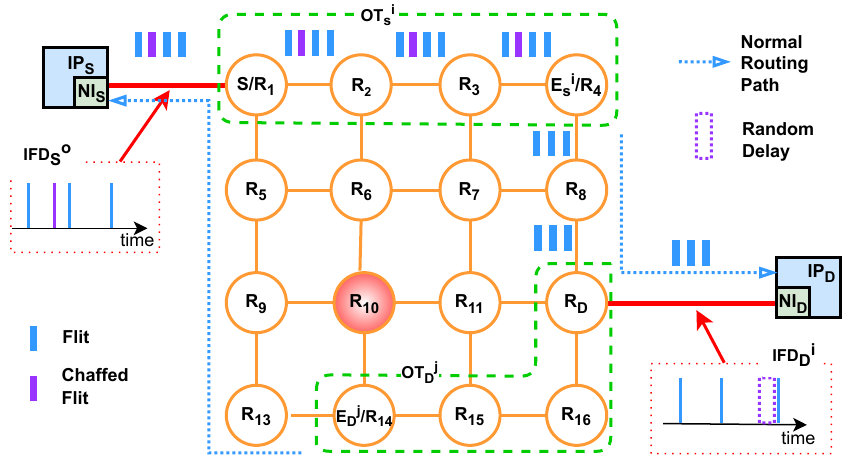}
\vspace{-0.1in}
\caption{
Two separate outbound tunnels $OT_S^i$ and $OT_D^j$ are used by $IP_S$ and $IP_D$ for communication. In $IP_S$ to $IP_D$ communication, chaffed flit is inserted at $NI_S$ and winnowed at $E^i_S$ ($R_4$). $E^i_S$ adds random delay to the flit sequence. The packet follows normal routing after an outbound tunnel ends.}
\label{fig:example_of_counter}
\vspace{-0.1in}
\end{figure}

An outbound tunnel ($OT^i_S$) is a route created from the source router ($S$) of the tunnel to an arbitrary router called \textit{tunnel endpoint} ($E^i_S$). Here, $i$ indicates the parameter for each tunnel instance. 
Figure~\ref{fig:example_of_counter} shows how outbound tunnels, $OT_S^i$ and $OT_D^i$, are used when $IP_S$ and $IP_D$ are injecting packets to the network.
It is important to highlight that these $OT^i$s are only bound to their source router and are independent of any communication session.
Each tunnel is associated with a timeout bound. After the timeout, the tunnel that belongs to a particular source $S$ will cease to exist and a new tunnel will be created with a different endpoint ($E^{i+1}_S$). $E_S^i$ of an $OT_S^i$ is randomly selected from any router that is $h_{min}$ to $h_{max}$ hops away from the source of the tunnel. We use $h_{min} = 3$ because a minimum of three nodes are needed for anonymous routing and increasing it further will negatively affect the performance~\cite{dingledine2004tor}. $h_{max}$ can be configured to balance the performance and the number of endpoints. 

\begin{figure}[t]
\includegraphics[width=0.9\linewidth]{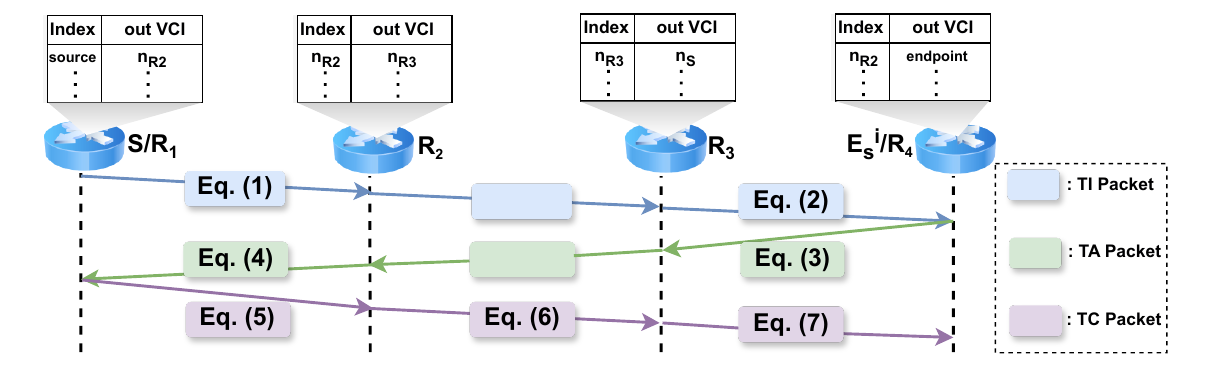}
\vspace{-0.1in}
\caption{ Message transfer in a three-way handshake to create an outbound tunnel between router $R_1$ and $R_4$ and final state routing tables of each router representing the outbound tunnel.}
\vspace{-0.1in}
\label{fig:tunnel_creation}
\end{figure}

Figure~\ref{fig:tunnel_creation} zooms into the tunnel creation phase. A summary of notations used in tunnel creation can be found in Table \ref{tab:notation}.
Tunnel creation is a three-way handshake process. 
The source broadcasts a Tunnel Initialization (TI) packet to all the routers and only $E_S^i$ responds back to the source with a Tunnel Acceptance (TA) packet. Once the source receives an $ACK$ from $E_S^i$, it sends the Tunnel Confirmation (TC) packet to $E_S^i$.
After these three steps, each router in the tunnel has two random Virtual Circuit Identifiers (VCI) saved in their routing table to define the succeeding and preceding hops representing the tunnel. For the rest of the section, we refer to $E^i_S$ as just $E$.

\begin{table}[b]
\caption{\small Notations used in tunnel creation.}
\vspace{-0.15in}
\begin{center}
\begin{tabular}{r p{7cm} }
\toprule
$E\hat{n}_K$ & Encrypts message $M$ using key $K$ \\
$D\hat{e}_K$ & Decrypts message $M$ using key $K$ \\
$OPuK_S^i$& One-time public key used by source $S$ \\
$OPrK_S^i$& Corresponding private key to $OPuK_S^i$ \\
$PuK_E$ & Global public key of $E$ \\
$PrK_E$ & Corresponding private key to $PuK_E$ \\
$TPuK_R^i$ & Temporary public key of node $R$ \\
$TPrK_R^i$ & Corresponding private key to $TPuK_R^i$ \\
$K_{S-R}$ & Symmetric key shared between $S$ and $R$ \\
$n_R$ & Random nonce generated by node $R$ \\
$r$ & Random number generated by $S$ \\
$pkt[i]$ & $i^{th}$ element of a packet $pkt$ \\
$pre(R)$ & Previous router (in upstream direction) \\
$next(R)$ & Next router (in downstream direction) \\
$rand(a,b)$ & Generates random number between $a$ and $b$ \\
\bottomrule
\end{tabular}
\label{tab:notation}
\end{center}
\vspace{-0.1in}
\end{table}

\subsubsection{Tunnel Initialization}
In the example (Figure.~\ref{fig:example_of_counter}), $S$ sends a TI packet as:
\vspace{-0.05in}
\begin{align}
\{TI || OPuK_S^{i} || E\hat{n}_{PuK_E}(OPuK_S^{i}||r)||TPuK_S^{i}\}
\vspace{-0.1in}
\end{align}
$TI$ identifies the packet as a Tunnel Initiation packet. $OPuK_S^{i}$ is the sources' one-time public key for the $i^{th}$ tunnel and $OPrK_S^i$ is the corresponding private key. In other words, an $OT^i_S$ can be uniquely identified by this key pair. $PuK_E$ and $PrK_E$ are the global public and private keys of $E$, respectively. They will not be changed with each tunnel creation. $OPuK_S^{i}$ and a randomly generated value $r$ is concatenated and encrypted through public-key encryption using the key $PuK_E$ ($E\hat{n}_{PuK_E}$).
Only $E$ can decrypt this encryption because only E has the corresponding private key ($PrK_E$). Finally, the temporary public key ($TPuK_S^{i}$) is concatenated at the end of the packet.
TI packet is broadcasted instead of directly routed to avoid anonymity being broken at its birth.

\begin{algorithm}[t]
  \caption{TI Packet handling at $R$}
  \label{algo:TI_packet_handling}
  \begin{algorithmic}[1]
    \State $pkt$ :A TI packet
    \Procedure{HandleTI }{$pkt$} 
        \If{ $OPuK_S^{i}$ in \textit{TL table}}
            \State discard $pkt$
           \label{algo_line:TI_packet_handling-line3}
        \Else
            \State store $OPuK_{S}^{i}$ and $TPuK_{pre(R)}^{i}$
            \label{algo_line:TI_packet_handling-line5}
        \EndIf
        \If{ $D\hat{e}_{PrK_R}(pkt[3])$ is successful} 
            \State G{\scshape enerate}TA ($D\hat{e}_{PrK_R}(pkt[3])$, $pkt[4]$)
            \label{algo_line:TI_packet_handling-line8}
        \Else
            \State $pkt[4] \gets TPuK_{R}^{i}$
            \label{algo_line:TI_packet_handling-line10}
            \State forward $pkt$
            \label{algo_line:TI_packet_handling-line11}
        \EndIf
    \EndProcedure
  \end{algorithmic}
\end{algorithm}

Any Router ($R$) receiving a TI packet will follow Algorithm~\ref{algo:TI_packet_handling}. Tunnel Lookup (TL)  table has unique entries for every TI packet comes to the router. First, it tries to match $OPuK_S^{i}$ with the existing entries in the TL table. On match, the message will get discarded to avoid any duplication due to TI packet broadcasting (line \ref{algo_line:TI_packet_handling-line3}). Otherwise, $OPuK_S^{i}$ and $TPuK_{pre(R)}^{i}$ are stored in the TL table (line \ref{algo_line:TI_packet_handling-line5}). Next, $R$ will try to decrypt the message and if it is successful, it should recognize itself as the intended endpoint and run Algorithm~\ref{algo:generate_TA_packet} (line \ref{algo_line:TI_packet_handling-line8}). If not, $R$ will replace $TPuK_{pre(R)}^{i}$ with its own temporary key $TPuK_{R}^{i}$ and forward the TI packet to the next hop ($next(R)$)(line \ref{algo_line:TI_packet_handling-line10} and \ref{algo_line:TI_packet_handling-line11}).
For example, in figure~\ref{fig:example_of_counter}, after receiving a TI packet from $R_2$, $R_3$ will generate and forward the following TI packet to $R_4$:
\vspace{-0.1in}
\begin{align}
\{TI || OPuK_S^{i} || E\hat{n}_{PuK_E}(OPuK_S^{i}||r)||TPuK_{R_3}^{i}\}
\vspace{-0.1in}
\end{align}

\subsubsection{Tunnel Acceptance}

\begin{algorithm}[t]
  \caption{TA packet generation at $E$}
  \label{algo:generate_TA_packet}
  \begin{algorithmic}[1]
    \State $parm_1$ : parameter resolved to $OPuK_S^{i}||r$
    \Procedure{GenerateTA }{$in_1$, $TPuK_R$} 
        \If{$parm_1[1] \neq OPuK_S^{i}$ } 
        \label{algo_line:generate_TA_packet-line2}
            \State discard the $pkt$
            \label{algo_line:generate_TA_packet-line3}
        \Else
            \State generate and store $n_E$ and $K_{S-E}$
            \label{algo_line:generate_TA_packet-line5}
            \State $enc \gets E\hat{n}_{TPK_{next(E)}^{i}}(E\hat{n}_{OPuK_{S}^{i}}(r|| n_E||$ $K_{S-E}))$
            \label{algo_line:generate_TA_packet-line6}
            \State \Return $\{TA || enc \}$
        \EndIf
    \EndProcedure
  \end{algorithmic}
\end{algorithm}

Upon receiving the TI packet, $E$ runs Algorithm~\ref{algo:TI_packet_handling} first and then calls Algorithm~\ref{algo:generate_TA_packet} as the endpoint of the tunnel. Algorithm \ref{algo:generate_TA_packet} shows the outline of TA packet generation at any endpoint router ($E$). First, $E$ validates the integrity of the packet by comparing decrypted $OPuK_S^{i}$ value and plaintext $OPuK_S^{i}$ value (line \ref{algo_line:generate_TA_packet-line2}). If the packet is validated for integrity, Algorithm \ref{algo:generate_TA_packet} will execute the following steps. First, it will generate random nonce $n_E$ which will be used as VCI. Next, it will generate a symmetric key $K_{S-E}$ to use between $S$ and $E$. Then it will log both $n_E$ and $K_{S-E}$ in the TL table and $n_E$ in the \textit{routing table} as indexed VCI (line \ref{algo_line:generate_TA_packet-line5}). Next, it will perform encryption of the concatenation of $n_E$, $K_{S-E}$ and $r$ using the key $OPuK_S^{i}$ which will allow only $S$ to decrypt the content (line \ref{algo_line:generate_TA_packet-line6}). Finally, the resultant encryption is encrypted again by the $TPuK_{next(E)}$ (line \ref{algo_line:generate_TA_packet-line6}).
In the figure~\ref{fig:example_of_counter}, $E^i_S$ will generate the following TA packet: 
\vspace{-0.05in}
\begin{align}
\{TA || E\hat{n}_{TPuK_{R_3}^{i}}(E\hat{n}_{OPuK_{S}^{i}}(r|| n_E||K_{S-E}))\}
\end{align}
\vspace{-0.2in}
\begin{algorithm}[b]
  \caption{TA packet handling at R}
  \label{algo:handling_TA_packet}
  \begin{algorithmic}[1]
  \State $pkt$ : A TA packet
    \Procedure{HandleTA }{$pkt$} 
        \If{ $R$ is $S$ of $OT^i$}
        \label{algo_line:handling_TA_packet-line2}
            \State G{\scshape enerate}TC ($pkt$)
            \label{algo_line:handling_TA_packet-line3}
        \Else
            \State $dct \gets D\hat{e}_{TPrK_{R}^{i}}(pkt[2])$ 
            \label{algo_line:handling_TA_packet-line5}
            \State generate and store $n_R$ and $K_{S-R}$
            \label{algo_line:handling_TA_packet-line6}
            \State $enc \gets E\hat{n}_{OPuK_{S}^{i}}(dct || n_R || K_{S-R})$
            \label{algo_line:handling_TA_packet-line7}
            \State $enc \gets E\hat{n}_{TPuK_{next(R)}^{i}}(enc)$
            \label{algo_line:handling_TA_packet-line8}
            \State \Return $\{TA || enc \}$
        \EndIf
    \EndProcedure
  \end{algorithmic}
\end{algorithm}

When a router $R$ receives a TA packet, it will execute Algorithm~\ref{algo:handling_TA_packet}. If the router is the source of the $OT^i$, it will execute Algorithm~\ref{algo:generating_TC_packet} (line \ref{algo_line:handling_TA_packet-line3}). Otherwise, it will go through the following steps. 
First, it decrypts the packet using the temporary private key ($TPrK_{R}^{i}$) (line \ref{algo_line:handling_TA_packet-line5}) and generates a random nonce and symmetric key ($n_{R}$, $K_{S-R}$). This generated $n_{R}$ and $K_{S-R}$ are stored in $R$'s TL table (line \ref{algo_line:handling_TA_packet-line6}).
The nonce and symmetric key pair is concatenated to the decrypted packet ($dct$) (line \ref{algo_line:handling_TA_packet-line6} and \ref{algo_line:handling_TA_packet-line7}), which will be encrypted using source public key ($OPuK_{S}^{i}$) to add another layer of security (line \ref{algo_line:handling_TA_packet-line7}). Finally, $R$ will encrypt the content with the public key of the next hop $next(R)$ (line \ref{algo_line:handling_TA_packet-line8}).
In the example, $R_2$ forwards the following TA packet to $R_1$:
\begin{align}
    \{TA || E\hat{n}_{TPuK_{R_2}^{i}}(E\hat{n}_{OPuK_{S}^{i}}(E\hat{n}_{OPuK_{S}^{i}}(E\hat{n}_{OPuK_{S}^{i}}( r||n_E||K_{S-E}) || n_{R_3}||K_{S-R_3} )||n_{R_2}||K_{S-R_2}))\}
\end{align}

\begin{algorithm}[t]
  \caption{TC packet generation at S}
  \label{algo:generating_TC_packet}
  \begin{algorithmic}[1]
    \State $pkt$ :A TA packet
    \Procedure{GenerateTC }{$pkt$} 
        \State $dec \gets D\hat{e}_{TPrK_{S}^{i}}(pkt[2])$ 
        \label{algo_line:generating_TC_packet-line2}
        \For{no of hops in OT}
        \label{algo_line:generating_TC_packet-line3}
            \State $dec \gets D\hat{e}_{OPrK_{S}^{i}}(dec)$
            \label{algo_line:generating_TC_packet-line4}
        \EndFor
        \label{algo_line:generating_TC_packet-line5}
        \State $enc \gets r$
        \For{R = E to next(S)}
        \label{algo_line:generating_TC_packet-line7}
            \State $enc \gets n(R) || E\hat{n}_{K_{S-R}}(enc)$
            \label{algo_line:generating_TC_packet-line8}
        \EndFor
        \label{algo_line:generating_TC_packet-line9}
        \State \Return $\{TC || enc \}$
        \label{algo_line:generating_TC_packet-line10}
    \EndProcedure
  \end{algorithmic}
\end{algorithm}

\subsubsection{Tunnel Confirmation}
Algorithm \ref{algo:generating_TC_packet} depicts the TC packet generation at the source router $S$. $TPrK_{S}^{i}$ is used to decrypt the outermost encryption (line \ref{algo_line:generating_TC_packet-line2}), then each layer of the inner encryption is peeled away using the $OPrK_{S}^{i}$ (loop from line \ref{algo_line:generating_TC_packet-line3} to \ref{algo_line:generating_TC_packet-line5}). $S$ extracts information of all the VCIs and symmetric keys. $r$ is used to check the authenticity of the packet received (make sure the TA is a packet from the actual endpoint $E$). Finally, starting from $E$ to $pre(S)$ (reverse order of routers in tunnel excluding $S$), $r$ is encrypted by the respective symmetric key and concatenated with the respective nonce iteratively (loop from line \ref{algo_line:generating_TC_packet-line7} to \ref{algo_line:generating_TC_packet-line9}). In the figure~\ref{fig:example_of_counter}, $S$ generates the TC packet structured as:
\vspace{-0.1in}
\begin{align}
\{TC|| n_{R_2} || E\hat{n}_{K_{S-R_2}}(n_{R_3} || E\hat{n}_{K_{S-R_3}}( 
n_{E} || E\hat{n}_{K_{S-E}}(r))) \}
\end{align}

$TC$ denotes the packet type. The packet is layered and encrypted using the symmetric keys distributed in the previous stage. Here, $n_{*}$ represents the outgoing VCI at each router from $S$ to $prev(E)$. For example, $n_{R_2}$ defines outgoing VCI of $S$ and $n_{R_3}$ defines outgoing VCI for $R_2$. After the TC packet is received by each node, it decrypts the outermost layer and stores the corresponding outgoing VCI value in the routing table indexed as incoming VCI. For example, when $R_2$ receives the packet it will decrypt the content using key $K_{S-R_2}$ and store the outgoing VCI as $n_{R_3}$ in the routing table indexed as $n_{R_2}$. Similarly, all the routers in between $S$ and $E$ will update the routing table entry corresponding to the tunnel. In the example, $R_2$ will send $TC$ packet to $R_3$ structured as follows:

\vspace{-0.2in}
\begin{align}
\{TC|| n_{R_3} || E\hat{n}_{K_{S-R_3}}(n_{E} || E\hat{n}_{K_{S-E}}(r)) \}
\end{align}

\noindent
Finally, $R_3$ will send the TC packet to $E$ ($R_4$) structured as:

\vspace{-0.1in}
\begin{align}
\{TC|| (n_{E} || E\hat{n}_{K_{S-E}}(r) \}
\end{align}

The packet transfers during the tunnel creation is inherently secure. This is because the tunnel creation phase ensures the transfer of only publicly available information as plaintext, while all sensitive data (including VCIs) are encrypted using symmetric or asymmetric methods. The only way an attacker can break anonymity of a packet is by knowing all the VCIs of the tunnel. This is not possible since the asymmetric and symmetric cryptography are computationally secure.

\subsection{Data Transfer}
A previously created outbound tunnel ($OT_S^i$) is used to transfer messages anonymously from $IP_S$ to $IP_D$. Before transferring the packet, the source will encrypt the actual destination header using the key $K_{S-E}$ which is the symmetric key shared between the source and endpoint during the tunnel creation. When we consider the data transfer through tunnel $OT_S^i$, a Data Transfer (DT) packet is injected into the tunnel by S structured as:
\vspace{-0.05in}
\begin{align}
\{DT || n_{R_2} || E\hat{n}_{K_{S-E}}(D) || E\hat{n}_D(M) \}
\end{align}

Here, DT is the packet type identifier, $n_{R_2}$ is the outgoing VCI, and $E\hat{n}_D(M)$ is the encrypted payload of the packet. At the router, $R_2$, the outgoing VCI is identified through a simple routing table lookup on the incoming VCI of the packet. Then $R_2$ replaces the outgoing VCI of the packet ($n_{R_2}$) with the next outgoing VCI, which is $n_{R_3}$, and routes the packet to the next hop. Similarly, $R_3$ replaces the outgoing VCI of the packet to $n_{E}$. Note that any intermediate node including $E$ does not know both source and destination of a single packet which ensures anonymity. 

\subsection{Traffic Obfuscation}
\label{subsec:obf-tech}

The main intuition behind the traffic obfuscation is to add noise to inbound and outbound flows ($IFD_S^o$, $IFD_D^i$), so it will be harder for ML-model to do accurate flow correlation. Section \ref{subsec:obf-tech-chaff} and \ref{subsec:obf-tech-delay}  introduce two traffic obfuscation techniques.

\subsubsection{Obfuscation with Chaffs}
\label{subsec:obf-tech-chaff}
 
We introduce a chaffing scheme as our first obfuscation technique. Chaff is a dummy flit with no usable data. Specifically, we insert chaff/chaffs in outbound tunnel traffic at the network interface of the source and filter out chaffs at the endpoint of the tunnel. The outbound flow ($IFD_S^o$) will have inter-flit delay data relevant to both chaffs and legitimate flits but inbound flow ($IFD_D^i$) will have inter-flit delay data relevant only to legitimate flits. Algorithm \ref{algo:adding_chaff} describes the chaffing process at the NI of the source. W{\scshape akeup} procedure (Line \ref{algo_line:adding_chaff-line1} - \ref{algo_line:adding_chaff-line4}) is the periodical function called by every NI in every clock cycle. We introduce a procedure named A{\scshape ddChaff} (Line \ref{algo_line:adding_chaff-line5} - \ref{algo_line:adding_chaff-line23}) to obfuscate traffic through chaffing. 

We insert chaffs in two specific scenarios to ensure the obfuscation scheme works with the majority of traffic patterns: (i) \textit{first scenario}: insert chaffs in the long gap between flits (line \ref{algo_line:adding_chaff-line6} - \ref{algo_line:adding_chaff-line13}), and (ii) \textit{second scenario}:  insert chaff flit in middle of closely packed flits (line \ref{algo_line:adding_chaff-line14} - \ref{algo_line:adding_chaff-line21}). When the outbound link of source NI is idle for more than $T_c$ cycles (Line \ref{algo_line:adding_chaff-line6}, the first scenario is considered. The intuition behind this method is to hinder the possibility of ML-model using long inter-flit delays of inbound and outbound flows in sparse traffic scenarios. In order to control overhead, we use a percentage threshold ($P_c$) and ensure only $P_c$ of idle gaps between packets get obfuscated (line \ref{algo_line:adding_chaff-line8} - \ref{algo_line:adding_chaff-line9}). If chosen to be obfuscated, a dummy packet of is created and is enqueued to the output queue of the NI (line \ref{algo_line:adding_chaff-line10} - \ref{algo_line:adding_chaff-line13}). At the endpoint of the tunnel, the dummy flits are filtered out and discarded. The \textit{chaffId} header is used to identify chaffed flit or packet. In the first scenario, a hash of the NI identification number is used as chaffed id. Unlike other headers, this header is encrypted by $K_{S-E}$ which is the symmetric key shared between the source ($S$) and endpoint ($E$) during the tunnel creation (line \ref{algo_line:adding_chaff-line12}). Therefore, only endpoint can filter chaffed packets by decrypting $E\hat{n}_{K_{S-E}}$(\textit{hash($NI_{ID}$)}. 

When the input queue of source NI received a packet (line \ref{algo_line:adding_chaff-line14}), the first scenario is considered (line \ref{algo_line:adding_chaff-line15} - \ref{algo_line:adding_chaff-line21}). The intuition behind this scenario is to hinder the possibility of ML-model using burst of small inter flits delays of inbound and outbound flows in heavy traffic. The example shown in Figure \ref{fig:example_of_counter} demonstrates the chaffing in second scenario and removing that chaff. Here, $P_c$ limits the number of packets being obfuscated (line \ref{algo_line:adding_chaff-line16} - \ref{algo_line:adding_chaff-line17}). If chosen to be obfuscated, chaff is inserted in the middle of legitimate packets at a random position (Line \ref{algo_line:adding_chaff-line18} - \ref{algo_line:adding_chaff-line21}). $K_{S-E}$ is used to encrypt \textit{chaffId}, represents the  position of chaff flit, which is used by the endpoint to filter the chaffed flit. 

A random number generator is already in the NI for cryptographic process. Therefore, the same generator is used for random number generation in line \ref{algo_line:adding_chaff-line8}, \ref{algo_line:adding_chaff-line10} and \ref{algo_line:adding_chaff-line16}. If the \textit{cflag} (line \ref{algo_line:adding_chaff-line3}, \ref{algo_line:adding_chaff-line6}, \ref{algo_line:adding_chaff-line7}, \ref{algo_line:adding_chaff-line15} and \ref{algo_line:adding_chaff-line23}) variable is true, it indicates the current gap between flits was already checked for insertion of packet. It is important to note that, (1) the dummy flits are added only when outbound link is idle, therefore, it has less impact on the program running on source IP, and (2) the dummy flits will only impact at most 3 internal links associated with the tunnel, therefore, it has less impact on the other traffic in the network. The scenario two inserts relatively less number of dummy flits and they will only impact at most 3 internal links. Experimental results in Section \ref{subsec:ovehead_count} validate that our obfuscation technique only results in negligible overhead. 

\begin{algorithm}[t]
  \caption{Add Chaff at source NI}
  \label{algo:adding_chaff}
  \begin{algorithmic}[1]
  \Procedure{W{\scshape akeup} }{ }
      \label{algo_line:adding_chaff-line1}
      \State .....
      \State \textit A{\scshape ddChaff}($cflag$)
      \label{algo_line:adding_chaff-line3}
      \State .....
      \label{algo_line:adding_chaff-line4}
  \EndProcedure
  \Procedure{AddChaff}{$cflag$}
    \label{algo_line:adding_chaff-line5}
        \If{ \textit{$cflag$} = $False$ and \textit{getIdleCy($link_o$) $> T_c$}}
        \label{algo_line:adding_chaff-line6}
        \State \textit{$cflag$ = True}
        \label{algo_line:adding_chaff-line7}
        \State \textit{$randNo$} $\gets$ $rand (0, 99)$ 
        \label{algo_line:adding_chaff-line8}
            \If{ $randNo$ $\leq P_c$ }
                \label{algo_line:adding_chaff-line9}
                \State $nFlits$ $\gets$ $rand (4, 5)$ 
                \label{algo_line:adding_chaff-line10}
                \State \textit{$dPkt$ $\gets$ $new$ $Packet(nFlits)$}
                \label{algo_line:adding_chaff-line11}
                \State $dPkt.chId$ $\gets E\hat{n}_{K_{S-E}}(hash(NI_{ID})$)
                \label{algo_line:adding_chaff-line12}
                \State $outputQueue.enque(dummyPkt)$
                \label{algo_line:adding_chaff-line13}
            \EndIf 
        \EndIf
        \If{$inputQueue.recivedPacket()$ = $True$ }
        \label{algo_line:adding_chaff-line14}
        \State \textit{$cflag$ = True}
        \label{algo_line:adding_chaff-line15}
        \State \textit{$randNo$} $\gets$ $rand (0, 99)$ 
        \label{algo_line:adding_chaff-line16}
            \If{ $randNo$ $\leq P_c$ }
            \label{algo_line:adding_chaff-line17}
                \State \textit{$chId$} $\gets$ $rand (0, len(inputPkt))$
                \label{algo_line:adding_chaff-line18}
               \State $dFlit$ = $new$ $flit()$
               \label{algo_line:adding_chaff-line19}
               \State $encChId$ $\gets E\hat{n}_{K_{S-E}}(hash(NI_{ID})|chId$)
               \label{algo_line:adding_chaff-line20}
               \State $pkt.insert(chId, encChId, dFlit) $
               \label{algo_line:adding_chaff-line21}
            \EndIf
        \EndIf
        \If{$outputQueue.sendPacket()$ = $True$ }
        \label{algo_line:adding_chaff-line22}
            \State \textit{$cflag$ = $False$}
            \label{algo_line:adding_chaff-line23}
        \EndIf
    \EndProcedure
  \end{algorithmic}
\end{algorithm}

\subsubsection{Obfuscation with Random Delay}
\label{subsec:obf-tech-delay}

The second obfuscation technique adds random delays to selected flits and tries to tamper with the timing aspect of the traffic flow. Flits belonging to only $P_d$ percentage of packets are subject to added delays. The tunnel endpoint is responsible for adding delays. Traveling through the rest of the hops the flit propagates the delay to the destination tampering with timing features of the inbound flow ($IFD_D^i$). Figure \ref{fig:example_of_counter} demonstrates the effect of added delay in traffic flows.
It is clear that chaffing and random delays obfuscate the actual traffic between source and destination. Both of these techniques can be used simultaneously or in a standalone manner depending on the requirement. Experimental results (Table~\ref{tab:conter_evaluation1} and~\ref{tab:conter_evaluation2}) show that both techniques effectively defend against ML-based flow correlation attacks.
\section{Experimental Evaluation}
\label{sec:exp-eval}

\begin{table}[t]
\caption{System and interconnect configuration}
\vspace{-0.15in}
\label{table:configTable}
\begin{center}
\begin{tabular}{|p{0.5\textwidth} | p{0.30\textwidth}|}
\hline

\textbf{Parameter} & \textbf{Details} \\
\hline
Processor configurations & X86, 2GHz \\
\hline
L1 I \& D cache & 64KB, 64KB (64B block size) \\
\hline
Coherency Protocol & MI\\
\hline
Topology & 8$\times$8 Mesh \\
\hline


Chaffing rate ($P_c$) and delay addition rate ($P_d$) & 50\% \\
\hline
\end{tabular}
\label{tab1}
\end{center}
\vspace{-0.1in}
\end{table}

We model our proposed ML-based attack and countermeasures on a cycle-accurate Gem5~\cite{gem5}, a multi-core simulator, with Garnet 2.0~\cite{garnet} for the interconnection network modeling.
We use a 64-core system and the detailed system configuration is given in Table~\ref{table:configTable}. 
Splash-2~\cite{sakalis2016splash} benchmark applications as well as multiple synthetic traffic patterns were used in evaluation.
We used Pytorch library to implement the proposed DNN architecture. First, we show the results of the flow correlation attack in existing anonymous routing (ARNoC~\cite{charles2020lightweighta} and SAR~\cite{sarihi2021securing, patooghy2023securing}). Later, we show the robustness of the proposed anonymous routing protocol to mitigate the attack. In order to evaluate the area and energy overhead of our approach against state-of-the-art anonymous routing, we implemented ARNoC and our approach in Verilog and synthesized both designs using Synopsys Design Compiler with 32nm Synopsys standard cell library.

\subsection{Data Collection}
\label{sec:ex-data-col}

This section demonstrates the data collection on Gem5 for the training of DNN. Although the input to DNN is in the same structure, the inherent differences in synthetic traffic and real benchmarks led us to two ways of collecting flow pairs for training.

\subsubsection{Synthetic Traffic}

We performed data collection using Uniform-Random synthetic traffic with the following modification. All IPs send packets to randomly selected IPs except two ($IP_S$ and $IP_D$). These two IPs are the correlated pair communicating in a session. From all the packets injected from the source IP ($IP_S$), only $p$ percent of packets are sent to the destination IP ($IP_D$), and the remaining packets ($(100-p)$\%) are sent to other nodes. For example, $p = 80\%$ means 80\% of the total outbound packets from $IP_S$ will have $IP_D$ as the destination, while the other 20\% can have any other IP except IP $IP_S$ and $IP_D$ as the destination. Note that this 20\% can be viewed as noise from the perspective of communication between $IP_S$ and $IP_D$. 
Here, traffic between correlated pair models concentrated point-to-point traffic between two nodes (e.g., processing core and memory controller). The random point-to-point traffic models other NoC traffic in a heterogeneous SoC other than cache coherence traffic such as monitoring and management traffic, inter-process communication and message passing between heterogeneous IP cores. This randomized traffic between uncorrelated pairs introduces uncontrolled noise to correlated traffic flow. Therefore, random point-to-point synthetic traffic models worst-case-scenario for flow correlation attack.


To make the dataset generic, for a single $p$ value, we conduct experiments covering all possible mapping of correlated pairs to NoC nodes, which are 8064 mappings (64$\times$63$\times$2). We consider four traffic distributions with $p$ value of 95\%, 90\%, 85\%, and 80\%. In other words, we consider four different noise levels (5\%, 10\%, 15\% and 20\%) for our data collection simulations.
The full dataset  for a certain $p$ value contains 24192 flow pairs (\{$IFD_S^o$, $IFD_D^i$\}) which consists of 8064 correlated traffic flow pairs and 16128 uncorrelated traffic flow pairs. Note that for each correlated flow pair, we selected two arbitrary uncorrelated flow pairs. To evaluate our countermeasures, when collecting obfuscated traffic, we kept both $P_c$ and $P_d$ at 50\% to ensure uniform distribution of obfuscation. When obfuscating traffic using added delay, we vary the delay between $1-5$ cycles because a higher delay may lead to unacceptable performance overhead. We collected three categories of data sets: one with chaffing only, one with random delay only, and one with applying both chaffing and delaying simultaneously.

\subsubsection{Real Traffic}
\label{data_collection_real_traffic}

Here, we collect response cache coherence traffic from memory controller to requester. This is done via filtering out using virtual network (vnet) used for memory response traffic to requester which is vnet 4. We consider five Splash-2 benchmark application pairs running on two processors ($P_1$ and $P_2$) where two memory controllers ($MC_1$ and $MC_2$) are serving memory requests. The benchmark pairs used are \{{\it fft}, {\it fmm}\}, \{{\it fmm}, {\it lu}\}, \{{\it lu}, {\it barnes}\}, \{{\it barnes}, {\it radix}\}, \{{\it radix}, {\it fft}\}, where the first benchmark runs on $P_1$ and the second runs on $P_2$. The selected benchmarks have the diversity to make the dataset generic (for example, \textit{fft} and \textit{radix} are significantly different ~\cite{bienia2008parsec}). The address space of the benchmark running in $P_1$ is mapped only to $MC_1$. Therefore, $P_1$ only talks with the $MC_1$, and they are the correlated pair. The address space of the benchmark running in $P_2$ is assigned to both $MC_1$ and $MC_2$ in a way that, the ratio between memory request received by $MC_1$ from $P_1$ to memory request received by $MC_1$ from $P_2$ to be $p:(100-p)$. This percentage $p$ is similar to that of synthetic traffic and $(100-p)\%$ is the noise. For example, when $p=85\%$, $MC_1$ serves $15\%$ packets to $P_2$ when it severs $85\%$ packets to $P_1$.

Similar to synthetic traffic, we considered four values for $p$ which are 95\%, 90\%, 85\%, and 80\%. For a single $p$ value and a single benchmark pair, we conducted experiments covering all possible mapping of correlated pairs to NoC nodes, which are 4032 mappings (64$\times$63). The $MC_2$ and $P_2$ were randomly chosen in all these mappings. The full dataset for a certain $p$ value and benchmark pair contains 16128 flow pairs (4032 correlated pairs and 12096 uncorrelated pairs). To evaluate our countermeasures, we collect obfuscated data similar to synthetic traffic. We automated data collection using the gem5 simulator with a shell script that simulates different benchmarks and application mappings. This process produces traffic traces from gem5 as textual logs.
We then developed a Python script to pre-process these traces into 2D numpy arrays of inter-flit delays, serving as input for the DNN.


\subsection{Hyperparameter Tuning}
\label{sec:hyper}

Hyperparameters are parameters set before training to improve model performance, such as learning rate and filter size.
We rigorously tested various hyperparameter combinations to achieve superior attack success rates.
The training process consists of 10-20 epochs with a consistent learning rate of 0.0001. We performed batch normalization and adjusted the batch size to 10 for the training set. As for convolution layers (C1 and C2 in Figure~\ref{fig:dnn}), the channel size is selected as $k_1 = 1000$ and $k_2 = 2000$, with $w_1 = 5$ and $w_2 = 30$, for C1 and C2, respectively. As for fully connected layers, sizes are selected as 3000, 800, and 100 for FC1, FC2, and FC3, respectively.

There are a lot of challenges in tuning since the finalized parameters reflect a trade-off between cost and effectiveness. First, the learning rate of the raining was reduced from 0.001 to 0.0001 which increases the training time but successfully avoids the Local Minima problem. Our decision to limit training to 10-20 epochs was primarily based on initial experiments that employed early stopping based on validation error. This approach consistently showed that performance stabilized within this range. Additionally, restricting the number of epochs to 10-20 served as a regularization technique to further mitigate the risk of overfitting. Batch size is also decreased from 50 to 10. In this way, fewer samples are provided for one iteration of training, but it improves the stability of training progress. Additionally, the selection of parameters for convolution layers properly addressed their responsibilities. As discussed in Section~\ref{subsec:dnn-arc}, C1 focuses on extracting rough relationships while C2 on advancing features. Therefore, C2 possesses two times of channels of C1, and a wider stride (30:5) to improve efficiency.

\subsection{Training and Testing}
\label{sec:ex-train-test}
In our study, we randomly divided a dataset of flow pairs for a specific configuration into a 2:1 ratio for the training and testing sets. Flow pairs were labeled as `1' for correlated and `0' for uncorrelated. We assessed our experiments using following four evaluation metrics. 




\begin{multicols}{2}
\begin{itemize}
   \item \textbf{Accuracy:} $\frac{tp+tn}{tp+tn+fp+fn}$
   \item \textbf{Recall:} $\frac{tp}{tp+fn}$
   \item \textbf{Precision:} $\frac{tp}{tp+fp}$
   \item \textbf{F1 Score:} $2\frac{Precision \cdot Recall}{Precision + Recall}$
\end{itemize}
\end{multicols}


Here, $tp$, $tn$, $fp$ and $fn$ represent true positive, true negative, false positive, and false negative, respectively. Intuitively, recall is a measure of a classifier's exactness, while precision is a measure of a classifier's completeness, and F1 score is the harmonic mean of recall and precision. The reason for utilizing these metrics comes from the limitation of accuracy. For imbalanced test cases (e.g., $> 90\%$ positive labels), a naive ML model which gives always-true output can reach $>90\%$ accuracy. 
The goal of the attacker is to identify correlating node pairs and launch complex attacks. Here, $fn$ is when an actual correlating pair is tagged as non-correlating by the DNN. $fp$ is when an actual non-correlating pair is tagged as correlating by the DNN. From an attacker's perspective, the negative impact of wasting time on launching an unsuccessful attack on $fp$ is relatively low compared to an attacker missing a chance to launch an attack due to a $fn$. Therefore, recall is the most critical metric compared to others when evaluating this flow correlation attack.

\subsection{ML-based Attack on Synthetic Traffic}
\label{sec:attack-results}

We evaluated the proposed attack for all four traffic distributions. The traffic injection rate was fixed to 0.01, and the IFD array size to 250. Table~\ref{tab:attack_evaluation} summarizes the results of the attack on ARNoC. All the considered traffic distributions show good metric numbers. We can see a minor reduction in performance with a reducing value of $p$. This is expected because of the increase in the number of uncorrelated packets in correlated flow pairs, making the correlation hard to detect. Even for the lowest traffic distribution of 80\% between two correlating pairs, the attacking DNN is able to identify correlated and uncorrelated flow pairs successfully with good metric values.

\begin{table}[tp]
\caption{Performance of ML-based attack on existing anonymous routing (ARNoC~\cite{charles2020lightweight}) for different traffic distributions.}
\vspace{-0.3in}
\begin{center}
\begin{tabular}{|c|c|c|c|c|}
\hline
\textbf{$p$} & \textbf{Accuracy}& \textbf{Recall} & \textbf{Precision} & \textbf{F1 Score} \\
\hline
95& 97.16\%& 91.98\%& 99.47\% & 95.58\%\\
\hline
90& 97.04\%& 93.35\%& 97.50\% & 95.38\%\\
\hline
85& 94.64\%& 91.32\%& 92.30\% & 91.81\%\\
\hline
80& 91.70\%& 80.02\%& 94.10\%& 86.60\%\\
\hline
\end{tabular}
\label{tab:attack_evaluation}
\end{center}
\vspace{-0.1in}
\end{table}

Table~\ref{tab:attack_evaluation_new_pro} summarizes the attack results on SAR, showing trends similar to those in the attack on ARNoC (Table~\ref{tab:attack_evaluation}). The main reason for this similarity is that our attack focuses on correlating inbound and outbound flows rather than focusing on breaking obfuscation techniques to hide communicating parties. Even though SAR uses packet-wise path diversity for anonymous routing, the proposed flow correlation attack performs well due to two reasons: (1) packet-level path diversity will not affect inter-flit delay inside the packet which is the fundamental feature of the proposed ML-based attack, and (2) since there are only three path scenarios (XY, YX, and XYX ) with a subtle variation, the variation of delay between flits of two adjacent packets is negligible to affect flow correlation adversely. These results confirm that our attack is realistic and can be applied on state-of-the-art anonymous routing (both ARNoC and SAR) to break anonymity across different traffic characteristics with varying noise. Due to the similarity of the attack performance in both anonymous routing protocols,  we only consider ARNoC for the subsequent experiments.



\begin{table}[tp]
\caption{Performance of ML-based attack on existing anonymous routing (SAR~\cite{sarihi2021securing,patooghy2023securing}) for different traffic distributions.}
\vspace{-0.3in}
\begin{center}
\begin{tabular}{|c|c|c|c|c|}
\hline
\textbf{$p$} & \textbf{Accuracy}& \textbf{Recall} & \textbf{Precision} & \textbf{F1 Score} \\
\hline
95& 96.91\%& 91.61\%& 99.07\% & 95.19\%\\
\hline
90& 96.67\%& 92.78\%& 96.90\% & 94.80\%\\
\hline
85& 94.59\%& 91.96\%& 91.53\% & 91.74\%\\
\hline
80& 92.30\%& 81.21\%& 94.97\%& 87.55\%\\
\hline
\end{tabular}
\label{tab:attack_evaluation_new_pro}
\end{center}
\vspace{-0.1in}
\end{table}

\subsection{Stability of ML-based Attack}
\label{sec:stability-ML-model}

In this section, we assess the stability of the proposed ML-based attack by varying configurable parameters. For experiments in this section, we use synthetic traffic with the value of $p$ as 85\% and the rest of the parameters as discussed in the experimental setup except for the varying parameter. 

\begin{table}[t]
\caption{Performance of ML-based attack on existing anonymous routing for different traffic injection rates.}
\vspace{-0.15in}
\begin{center}
\begin{tabular}{|c|c|c|c|c|}
\hline
\textbf{TIR} & \textbf{Accuracy}& \textbf{Recall} & \textbf{Precision} & \textbf{F1 Score} \\
\hline
0.001& 95.32\%& 92.29\%& 93.51\% & 92.89\%\\
\hline
0.005& 94.72\%& 90.14\%& 93.98\% & 92.02\%\\
\hline
0.01& 94.64\%& 91.32\%& 92.30\% & 91.81\%\\
\hline
0.05& 93.86\%& 88.56\%& 92.67\%& 90.56\%\\
\hline
\end{tabular}
\label{tab:attack_evaluation_for_varying_tir}
\end{center}
\vspace{-0.1in}
\end{table}

\subsubsection{Varying traffic injection rates (TIR)}

We collected traffic data for four traffic injection rates: 0.001, 0.005, 0.01 and 0.05, and conducted the attack. Table~\ref{tab:attack_evaluation_for_varying_tir} provides detailed results on metrics over selected values. We can see a small reduction in overall metrics including recall, with the increase in injection rate. This is because, higher injection rates will create more congestion and buffering delays on NoC traffic. The indirect noise from congestion and buffering delays makes it slightly hard for the ML model to do flow correlation. Overall, our proposed ML model performs well in different injection rates since all the metrics show good performance.


\subsubsection{Varying IFD Array Size}

We collected traffic data by varying the size of IFD array size ($l$) in the range of 50 to 550 and conducted the attack on existing anonymous routing. Table~\ref{tab:attack_evaluation_for_varying_no_flits} shows detailed results on metrics over selected values. For a lower number of flits, the relative values of the recall and other metrics are low. However, with the increasing number of flits, the accuracy also improves until the length is 250. This is due to the increase in the length of the IFD array the ML model has more features for the flow correlation. After the value of 250, the accuracy saturates at around 94.5\%. In subsequent experiments, we kept $l$ to 250 because ML-based attack performs relatively well with less monitoring time.

\begin{table}[ht]
\caption{Performance metrics of ML-based attack on existing anonymous routing for varying number of flits.}
\vspace{-0.1in}
\begin{center}
\begin{tabular}{|c|c|c|c|c|}
\hline
\textbf{IFD Array size(\textit{l})} & \textbf{Accuracy}& \textbf{Recall} & \textbf{Precision} & \textbf{F1 Score} \\
\hline
50& 83.53\%& 96.45\%& 67.96\% & 79.74\%\\
\hline
100& 90.92\%& 96.17\%& 80.28\% & 87.51\%\\
\hline
150& 90.93\%& 74.10\%& 98.32\% & 84.51\%\\
\hline
250& 94.64\%& 91.32\%& 92.30\%& 91.81\%\\
\hline
350& 94.71\%& 86.21\%& 97.39\% & 91.46\%\\
\hline
450& 94.58\%& 93.21\%& 90.58\% & 91.87\%\\
\hline
550& 94.66\%& 88.97\%& 94.30\%& 91.56\%\\
\hline
\end{tabular}
\label{tab:attack_evaluation_for_varying_no_flits}
\end{center}
\vspace{-0.1in}
\end{table}

\begin{table}[tp]
\caption{Performance of ML-based attack on existing anonymous routing for different mesh sizes.}
\vspace{-0.15in}
\begin{center}
\begin{tabular}{|c|c|c|c|c|}
\hline
\textbf{Mesh Size} & \textbf{Accuracy}& \textbf{Recall} & \textbf{Precision} & \textbf{F1 Score} \\
\hline
4$\times$4& 94.76\%& 91.86\%& 92.63\% & 92.24\%\\
\hline
8$\times$8& 94.64\%& 91.32\%& 92.30\%& 91.81\%\\
\hline
16$\times$16& 92.72\%& 80.28\%& 96.98\% & 87.84\%\\
\hline
\end{tabular}
\label{tab:attack_evaluation_for_varying_mesh}
\end{center}
\end{table}

\subsubsection{Varying Network Size}

To evaluate the stability of the ML model on varying network sizes, we analyzed the model on 16 core system with 4$\times$4, 64 core system with 8$\times$8, and 256 core system with 16$\times$16 mesh topology. Table~\ref{tab:attack_evaluation_for_varying_mesh} shows the performance results of the ML model for different network sizes. Attack on 4$\times$4 mesh shows slightly good metric values compared to 8$\times$8. The attack on a 16$\times$16 network shows relatively low accuracy and recall, due to the network's larger size, which alters the temporal characteristics of the traffic. With four times as many nodes and roughly double the average hops (10.67 compared to 5.33 in the 8x8 mesh), the 16x16 mesh experiences more congestion. These conditions introduce additional noise, such as queuing delays, which affect the communication patterns observable through inter-flit delays. Despite these challenges, the achieved recall of 80.28\% is sufficiently high for a successful attack. Considering good accuracy and other metrics, our ML-based attack shows stability across different mesh sizes.




\color{black}

\subsection{ML-based Attack on Real Benchmarks}
\label{sec:real-attack-results}

We trained and tested the model using two techniques. In the first technique, we merge datasets of a single $p$ value across all 5 benchmark combinations outlined in Section~\ref{sec:ex-data-col} to create the total dataset. Therefore, the total dataset has 80640 flow pairs before the 2:1 test to train split. Table~\ref{tab:attack_evaluation_real} summarizes the results for the first technique across all $p$ values. Good metric numbers across all traffic distributions show the generality of the model across different benchmarks. In other words, our attack works well across multiple benchmarks simultaneously. Even 20\% noise ($p = 80$) shows recall value just remains robust around 98\%. While lower precision may lead to resources being spent on false positives, this issue is relatively minor compared to the potential harm posed by high recall rates. From an adversarial perspective, recall is the critical metric, as discussed in Section~\ref{sec:ex-train-test}.

\begin{table}[tp]
\caption{Performance of ML-based attack on ARNoC~\cite{charles2020lightweight} using MI protocol for different noise levels in real benchmarks.}
\vspace{-0.3in}
\begin{center}
\begin{tabular}{|c|c|c|c|c|}
\hline
\textbf{$p$} & \textbf{Accuracy}& \textbf{Recall} & \textbf{Precision} & \textbf{F1 Score} \\
\hline
95& 99.43\%& 99.79\%& 98.01\% & 98.89\%\\
\hline
90& 99.11\%& 99.84\%& 96.75\% & 98.27\%\\
\hline
85& 98.76\%& 98.16\%& 97.01\% & 97.58\%\\
\hline
80& 96.08\%& 98.79\%& 87.15\%& 92.61\%\\
\hline
\end{tabular}
\label{tab:attack_evaluation_real}
\end{center}
\end{table}


To evaluate our attack success across cache-coherence protocols, we trained and test model using MOESI-hammer protocol using first technique. To enable a fair comparison with previous MI protocol experiments, we kept the MOESI protocol private cache size for each node the same as that of the MI protocol. Specifically, we kept L1 instruction and data cache size 32KB and L2 cache size 64KB per node. Table~\ref{tab:attack_evaluation_real_MOESI} summarizes the attack results on systems with MOESI-hammer cache coherence protocol. Since we only focus on first $l = 450$ inter-flit delays, the less cache coherence traffic of MOESI-hammer protocol does not affect the training of the ML model. Comparable results across all traffic distributions similar to MI protocol demonstrate that our attack is successful across multiple cache coherence protocols. Therefore, for simplicity, we only consider MI cache coherence protocol for experiments involving real traffic in the remainder of this paper.

\begin{table}[tp]
\caption{Performance of ML-based attack on ARNoC~\cite{charles2020lightweight} using MOESI-hammer protocol across different traffic distributions with real benchmarks}
\vspace{-0.15in}
\begin{center}
\begin{tabular}{|c|c|c|c|c|}
\hline
\textbf{$p$} & \textbf{Accuracy}& \textbf{Recall} & \textbf{Precision} & \textbf{F1 Score} \\
\hline
95& 99.46\%& 99.85\%& 98.08\% & 98.96\%\\
\hline
90& 99.07\%& 99.81\%& 97.61\% & 98.18\%\\
\hline
85& 98.81\%& 98.45\%& 96.94\% & 97.69\%\\
\hline
80& 96.45\%& 98.81\%& 88.48\%& 93.36\%\\
\hline
\end{tabular}
\label{tab:attack_evaluation_real_MOESI}
\end{center}
\end{table}

\color{black}


When we compare the performance of attack on real traffic against synthetic traffic (Table \ref{tab:attack_evaluation}), attack on real traffic shows better performance. This is primarily for two reasons. (a) The synthetic traffic generation is totally random. More precisely, the interval between two packets is random and the next destination of a specific source is random. This level of randomness is not found in real traffic making flow correlation in real traffic relatively easy. (b) In synthetic traffic all 64 nodes talk with each other making higher buffering delays eventually making flow correlation harder. However, buffering delays have a minor impact compared to randomness. The second technique uses dataset of a single $p$ value and single benchmark pair. Table~\ref{tab:attack_evaluation_real_bench} summarizes the results for the second technique when $p=85$ across five benchmark pairs. All benchmarks display strong metrics, though accuracy and recall slightly decrease in the 3rd and 4th rows. Both benchmark pairs have \textit{barnes} benchmark, which has lowest bytes per instruction in all benchmarks \cite{woo1995splash}. This results in sparse inter-flit array, eventually making it relatively harder to do flow correlation.

Misclassifications can have significant implications, and it is important to consider them from the perspective of an attacker. Misclassifications can be divided into two types: false positives and false negatives. False positives occur when uncorrelated traffic is incorrectly identified as correlated. In this scenario, an adversary would wastefully allocate resources to act upon these false leads to launch further attacks, ultimately yielding no actual threat. While pursuing false leads might seem inefficient, adversaries usually have sufficient resources and can inflict significant damage when they correctly identify correlated nodes. This potential for harm outweighs the minor setbacks caused by occasional false positives. On the other hand, false negatives represent a critical error from the attackers standpoint. This type of error occurs when actual correlated communicating pair go undetected. Missing such opportunities can be detrimental to the adversary’s objectives, particularly if the goal is to cause maximum disruption.

\begin{table}[tp]
\caption{Performance of ML-based attack on ARNoC~\cite{charles2020lightweight} using when p=85  across real benchmark combinations.}
\vspace{-0.3in}
\begin{center}
\begin{tabular}{|c|c|c|c|c|}
\hline
\textbf{benchmark} & \textbf{Accuracy}& \textbf{Recall} & \textbf{Precision} & \textbf{F1 Score} \\
\hline
\{fft, fmm\}& 98.29\%& 99.11\%& 94.42\% & 96.71\%\\
\hline
\{fmm, lu\}& 99.32\%& 97.63\%& 99.70\% & 98.65\%\\
\hline
\{lu, barnes\}& 97.84\%& 91.60\%& 99.76\% & 95.51\%\\
\hline
\{barnes, radix\}& 96.14\%& 84.59\%& 99.73\%& 91.54\%\\
\hline
\{radix, fft\}& 96.62\%& 97.05\%& 90.66\%& 93.75\%\\
\hline
\end{tabular}
\label{tab:attack_evaluation_real_bench}
\end{center}
\vspace{-0.1in}
\end{table}

\subsection{Robustness of the Proposed Countermeasure}
\label{subsec:robustness_of_counter}

\begin{table*}[t]
    \caption{Performance metrics of ML-based attack on proposed lightweight anonymous routing for different traffic distributions when trained with non-obfuscated traffic and tested with obfuscated traffic}
   \vspace{-0.1in}    
  \centering
  \resizebox{\textwidth}{!}{%
  \begin{tabular}{|c||c|c|c|c||c|c|c|c||c|c|c|c|}
    \hline
    &
    \multicolumn{4}{|c||}{\textbf{Chaffing }} &                                           
    \multicolumn{4}{|c||}{\textbf{Delay}} &
    \multicolumn{4}{|c|}{\textbf{Chaffing + Delay }} \\
    \hline
    \textbf{$p$} & \textbf{Acc.}& \textbf{Rec.} & \textbf{Prec.} & \textbf{F1.} &
    \textbf{Acc.}& \textbf{Rec.} & \textbf{Prec.} & \textbf{F1.} &
    \textbf{Acc.}& \textbf{Rec.} & \textbf{Prec.} & \textbf{F1.} \\
    \hline
    95& 66.55\%& 0.3\%& 33.33\% & 0.6\% &
    81.32\%& 56.70\%& 81.67\% & 66.93\% &
    63.36\%& 14.37\%& 37.18\%& 20.70\%\\
    \hline
    90& 66.59\%& 25.68\%& 49.78\% & 33.88\% &
    71.73\%& 66.15\%&  56.48\%& 60.94\% &
     56.47\%& 42.27\%& 40.45\%& 41.34\%\\
    \hline
    85& 61.2\%& 2.6\%& 12.4\% & 4.4\% &
    72.57\%& 50.59\%& 60.61\%& 55.15\% &
    66.33\%& 40.50\%& 49.39\%& 44.51\% \\
    \hline
    80& 72.76\%& 26\%& 77.16\%& 38.89\% &
    73.41\%& 34.97\%& 70.37\%& 46.72\% &
    60.34\%& 30.72\%& 39.75\%& 34.65\%\\
    \hline
  \end{tabular}
  }
  \label{tab:conter_evaluation1}
\end{table*}

\begin{table*}[t]
  \caption{Performance metrics of ML-based attack on proposed lightweight anonymous routing for different traffic distributions when trained and tested with non-obfuscated traffic}
   \vspace{-0.1in}
  \centering
   \resizebox{\textwidth}{!}{%
  \begin{tabular}{|c||c|c|c|c||c|c|c|c||c|c|c|c|}
    \hline
    &
    \multicolumn{4}{|c||}{\textbf{Chaffing }} &                                           
    \multicolumn{4}{|c||}{\textbf{Delay}} &
    \multicolumn{4}{|c|}{\textbf{Chaffing + Delay }} \\
    \hline
    \textbf{$p$} & \textbf{Acc.}& \textbf{Rec.} & \textbf{Prec.} & \textbf{F1.} &
    \textbf{Acc.}& \textbf{Rec.} & \textbf{Prec.} & \textbf{F1.} &
    \textbf{Acc.}& \textbf{Rec.} & \textbf{Prec.} & \textbf{F1.} \\
    \hline
    95& 76.64\%& 33.60\%& 84.67\% & 48.11\% &
    94.22\%& 87.31\%& 94.99\% & 90.99\% &
    73.80\%& 25.49\%&  80.70\%& 38.75\% \\
    \hline
    90& 79.45\%& 43.71\%& 87.39\% & 58.28\% &
    93.58\%& 93.42\%& 87.66\% & 90.45\% &
    77.95\%& 46.9\%& 78.14\%& 58.69\%\\
    \hline
    85& 78.75\%& 38.93\%& 93.16\% & 54.92\% &
    90.65\%& 86.83\%& 84.99\%& 85.90\% &
    77.06\%& 48.85\%& 74.65\%& 59.05\%\\
    \hline
    80& 79.75\%& 74.41\%& 67.58\%& 70.83\% &
    87.70\%& 80.32\%& 82.08\%& 81.19\% &
    77.56\%& 74.41\%& 64.13\%& 68.89\% \\
    \hline
  \end{tabular}
  }
  \label{tab:conter_evaluation2}
  \vspace{-0.1in}
\end{table*}


We evaluate the robustness of our lightweight anonymous routing in two ways. First, we assess our countermeasure (Section~\ref{sec:countermeasure}) against the ML-based attack (Section~\ref{sec:attack}) on synthetic and real traffic. Second, we examine the overall effectiveness of our attack in breaking anonymity.

\begin{table}[t]
\caption{Performance metrics of ML-based attack on proposed lightweight anonymous routing different for noise levels on real benchmarks.}
\vspace{-0.1in}
\begin{center}
\begin{tabular}{|c|c|c|c|c|}
\hline
\textbf{$p$} & \textbf{Accuracy}& \textbf{Recall} & \textbf{Precision} & \textbf{F1 Score} \\
\hline
95 & 89.93\%	& 76.67\%	& 82.44\%	& 79.45\% \\
\hline
90 & 88.78\%	& 77.08\%	& 78.30\%	& 77.69\% \\
\hline
85 & 87.63\%	& 62.04\%	& 84.34\%	& 71.49\% \\
\hline
80 & 85.89\%	& 76.85\%	& 70.23\% &	73.39\% \\
\hline
\end{tabular}
\label{tab:counter_evaluation_real}
\vspace{-0.1in}
\end{center}
\end{table}

\begin{table}[t]
\caption{Performance metrics of ML-based attack on proposed lightweight anonymous routing for real benchmarks.}
\vspace{-0.3in}
\begin{center}
\begin{tabular}{|c|c|c|c|c|}
\hline
\textbf{benchmark} & \textbf{Accuracy}& \textbf{Recall} & \textbf{Precision} & \textbf{F1 Score} \\
\hline
 \{fft, fmm\}& 87.64\%& 65.91\%& 80.24\% & 72.37\%\\
\hline
\{fmm, lu\}& 90.67\%& 83.03\%& 80.20\% & 81.59\%\\
\hline
\{lu, barnes\}& 84.98\%& 57.75\%& 76.19\% & 65.70\%\\
\hline
\{barnes, radix\}& 84.24\%& 40.35\%& 97.59\%& 57.10\%\\
\hline
\{radix, fft\}& 82.60\%& 37.67\%& 82.66\%& 51.79\%\\
\hline
\end{tabular}
\label{tab:counter_evaluation_real_bench}
\end{center}
\vspace{-0.1in}
\end{table}

We evaluate our countermeasure against ML-based attacks in three configurations for synthetic traffic: (1) using chaffing, (2) using a delay, and (3) using both chaffing and delay to obfuscate traffic. For each of the three configurations, we evaluate the ML-based attack on two scenarios: (1) train with non-obfuscated traffic and test with obfuscated traffic (Table~\ref{tab:conter_evaluation1}), and (2) train and test with obfuscated traffic (Table~\ref{tab:conter_evaluation2}). In all three configurations, the attack on the first scenario has performed poorly (the proposed countermeasure defends very well). This is expected because the attacking DNN has not seen any obfuscated data in the training phase. If we focus on the scenario of using a delay to obfuscate traffic (Table~\ref{tab:conter_evaluation2}), we can see a significant reduction in all the metrics. Large drops in recall when using chaffing as the obfuscation technique validate that the proposed countermeasure produces a significant negative impact on attackers' end goals. Adding random delay reduces accuracy and recall by about 3\% compared to non-obfuscated traffic in all the traffic distributions. Whereas, combining chaffing with delay reduces accuracy and recall by about 3\% as compared to chaffing alone. In other words, combining two obfuscation techniques did not seem to have any synergistic effect. We recommend chaffing as a good obfuscation configuration since adding delay has only a small advantage despite its overhead. Note that the poor performance of added random delay as a countermeasure validates the fact that our proposed attack is robust against inherent random network delays in the SoC.

When evaluating the performance of countermeasures using benchmark applications, we consider only chaffing to obfuscate traffic. Furthermore, we only train and test with obfuscated traffic which guarantees to give a strong evaluation of the countermeasure. As discussed in section~\ref{sec:real-attack-results} we evaluate the countermeasure using two techniques, (1) merged datasets across benchmarks (Table~\ref{tab:counter_evaluation_real}) and (2) datasets per benchmark when $p$ value is fixed (Table~\ref{tab:counter_evaluation_real_bench}). 
When we focus on Table~\ref{tab:counter_evaluation_real}, we see an overall reduction of metric values compared to the attack without countermeasure. Even though the accuracy reduction is around 10\%, the countermeasure has reduced recall value drastically. This will negatively affect the attacker due to missing a chance to launch an attack due to higher $fn$. When we compare the performance of the countermeasure on real traffic against synthetic traffic (Table~\ref{tab:conter_evaluation2}), the countermeasure on synthetic traffic has performed relatively better. This is due to the same two reasons mentioned in section~\ref{sec:real-attack-results} briefly, the randomness of synthetic traffic and increased buffer delay because every node communicates.

We evaluate the anonymity of proposed lightweight anonymous routing in three attacking scenarios. The first scenario is when \textit{one of the intermediate routers in the outbound tunnel is malicious}. The malicious router only knows the identity of the preceding and succeeding router, so the anonymity of the flits traveling through the tunnel is secured. The second scenario is when \textit{the tunnel endpoint is malicious}. The router will have the actual destination of the packet but not the source information; therefore by having a single packet, the malicious router cannot break the anonymity. This scenario is also considered secure in the traditional onion routing threat model~\cite{dingledine2004tor}. Complex attacks in malicious routers need a considerable number of packets/flits to be collected. It is hard due to two following reasons: (1) Our proposed solution changes the outbound tunnel of a particular source frequently. (2) Since the source and destination have two independent outbound tunnels, it is infeasible to collect and map request/reply packets.
The final scenario is when \textit{an intermediate router in a normal routing path is malicious}. This scenario arises when flits use normal routing after it comes out of the outbound tunnel. Similar to the previous scenario, the packet only knows about the true destination, and anonymity is not broken using a single packet. In other words, outbound tunnels change frequently, and the source and destination have different tunnels making it hard to launch complex attacks to break anonymity by collecting packets.

The robustness of our approach can be evaluated in terms of deadlock handling. We have implemented our model using Garnet 2.0, where the XY routing mechanism is used to guarantee deadlock-free communication. When we focus on our countermeasure, the first step of tunnel creation (Tunnel Initialization) uses the existing XY routing protocol to broadcast TI packets. The path of the TI packet determines the tunnel shape. Since a TI packet cannot take a Y to X turn, any tunnel created on XY routing inherently uses only XY turns inside the tunnel. Hence, in the data transfer phase, all the communication inside and outside the outbound tunnel will only take X to Y turns, ensuring deadlock-free communication.  

\subsection{Overhead of the Proposed Countermeasure}
\label{subsec:ovehead_count}

\begin{figure}[t]
\begin{subfigure}[b]{0.20\textwidth}
         \includegraphics[width=1.5\linewidth]{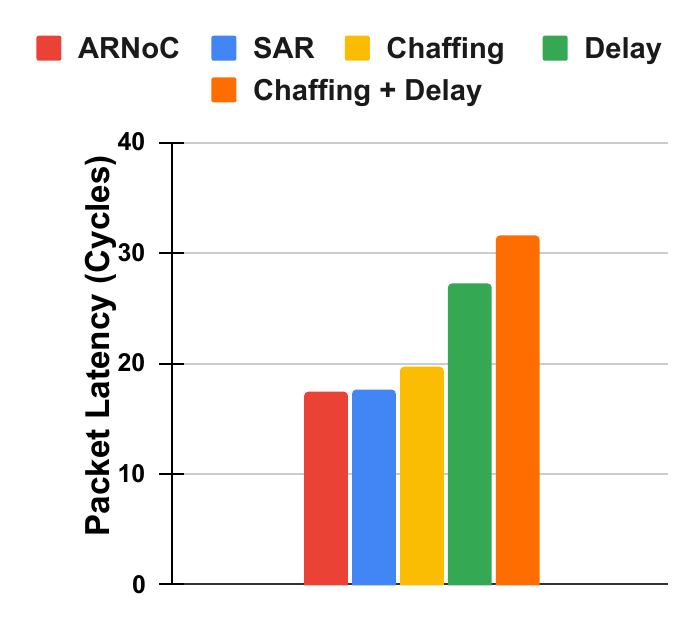}
         \vspace{-0.3in}
         \caption{Packet latency}
        \label{fig:latency}
\end{subfigure}
\hspace{1.2in}
\begin{subfigure}[b]{0.20\textwidth}
         \includegraphics[width=1.5\linewidth]{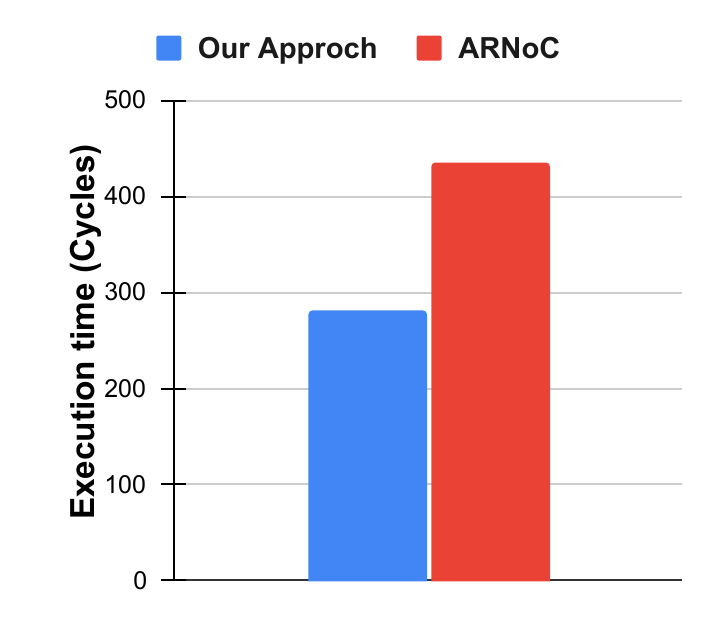}
         \vspace{-0.3in}
         \caption{Execution time}
        \label{fig:time}
\end{subfigure}
\vspace{-0.15in} 
\caption{Comparison of proposed countermeasure versus ARNoC and SAR: (a) average packet latency of data transfer, and (b) average execution time for tunnel creation (SAR does not have tunnel creation phase).}
\vspace{-0.1in} 
\label{fig:res23}
\end{figure}

Figure \ref{fig:latency} shows the average packet latency for our proposed lightweight countermeasure over ARNoC~\cite{charles2020lightweight} and SAR~\cite{sarihi2021securing,patooghy2023securing} in the data transmission phase. Obfuscating with chaff flit, which is the recommended obfuscation technique from Section~\ref{subsec:robustness_of_counter}, has only a 13\% and 11\% increase in performance overhead compared to ARNoC and ~\cite{sarihi2021securing}, respectively. Our approach reduces tunnel creation overhead by 35.53\% compared to ARNoC, as shown in Figure \ref{fig:time}, due to our strategy of creating shorter, outbound-only tunnels from the source to the random router (tunnel endpoint), unlike ARNoC's longer source-to-destination tunnel for outbound and inbound traffic. SAR does not have a tunnel creation phase. A key aspect of our approach is that tunnel creation occurs in the background, ensuring it does not directly impact data transfer performance. Overall, our approach is lightweight compared to ARNoC and has negligible performance overhead against~\cite{sarihi2021securing} while delivering both packet-level and flow-level anonymity.

\begin{table}[tp]
\caption{Comparison of area and energy overhead between ARNoC and proposed countermeasure in NoC.}
\vspace{-0.15in}
\begin{center}
\begin{tabular}{|c|c|c|c|}
\hline
& \textbf{ARNoC}& \textbf{Our Approach}& \textbf{Overhead} \\
\hline
Area($\mu m^2$) & 2914300 & 2957429 & + 1.47\% \\
\hline
Energy($mJ$)& 54.04 & 55.45 & + 2.6\% \\
\hline
\end{tabular}
\label{tab:area_and_energy_comparisson}
\end{center}
\vspace{-0.1in}
\end{table}

In addition to low-performance overhead, our lightweight anonymous routing has the inherent advantage of utilizing any adaptive routing mechanisms supported by NoC architectures (endpoint of output tunnel to the destination), while ARNoC cannot accommodate adaptive routing protocols because of having a pre-built tunnel from the source to destination. Similarly, SAR cannot accommodate adaptive routing due to its anonymous routing solution tightly bound to XY, YX, and XYX routing patterns.  Table \ref{tab:area_and_energy_comparisson} compares the area and energy overhead of our lightweight countermeasure against ARNoC in 8$\times$8 mesh topology. In the implementation, our approach uses only the chaffing obfuscation. The energy consumption was calculated by averaging the energy consumption of running the FFT benchmark across all possible mappings of the processing node and memory controller in mesh NoC-based SoC as discussed in Section~\ref{data_collection_real_traffic}. We observe a 1.47\% increase in area and a 2.6\% increase in energy. The area and energy overhead are negligible considering the performance improvement and additional security provided by our proposed anonymous routing compared to the state-of-the-art anonymous routing ARNoC. 


\color{black}



\section{Conclusion}
\label{sec:conclusion}

Network-on-Chip (NoC) is a widely used solution for on-chip communication between Intellectual Property (IP) cores in System-on-Chip (SoC) architectures. Anonymity is a critical requirement for designing secure and trustworthy NoCs. In this paper, we made two important contributions. We proposed a machine learning-based attack that uses traffic correlation to break the state-of-the-art anonymous routing for NoC architectures. We developed a lightweight and robust anonymous routing protocol to defend against ML-based attacks. Unlike existing anonymous routing protocols that only offer anonymity at the packet level, our proposed protocol enhances security by providing anonymity at both the packet level and the flow level.   Extensive evaluation using real as well as synthetic traffic demonstrated that our ML-based attack can break anonymity with high accuracy (up to 99\%) for diverse traffic patterns. The results also reveal that our lightweight anonymous routing protocol that uses chaffing as traffic obfuscation is robust against ML-based flow correlation attacks with minor performance and hardware overhead.


\section*{Acknowledgments}
This work was partially supported by National Science Foundation (NSF) grant SaTC-1936040.


\bibliographystyle{ACM-Reference-Format}
\bibliography{bibliography}


\end{document}